\documentclass[12pt]{article}
\usepackage{amsthm,amsmath,amsfonts,amssymb,cases}
\usepackage{calc} 
\usepackage{comment}
\usepackage{enumerate}
\usepackage{tikz}
\usepackage{pgfplots}
\usepackage[shortlabels]{enumitem}
\usepackage{hyperref}
\usepackage{cleveref}
\usepackage{float}
\usepackage{cite}

\newcommand{\R}{{\mathord{\mathbb R}}}
\newcommand{\Z}{{\mathord{\mathbb Z}}}
\newcommand{\N}{{\mathord{\mathbb N}}}
\newcommand{\C}{{\mathord{\mathbb C}}}

\newcommand{\HS}{{\mathord{\mathbb H}}}
\newcommand{\E}{{\mathord{\bf E}}}

\newcommand{\muu}[1]{\begin{align*} #1 \end{align*}}
\newcommand{\muun}[1]{\begin{align} #1 \end{align}}
\newcommand{\tcr}[1]{\textcolor{red}{#1}}
\newcommand{\SP}[1]{\left\langle #1  \right\rangle }
\renewcommand{\epsilon}{\varepsilon}

\newcommand{\vertiii}[1]{{\left\vert\kern-0.25ex\left\vert\kern-0.25ex\left\vert #1 
    \right\vert\kern-0.25ex\right\vert\kern-0.25ex\right\vert}}

\newcommand{\ben}{\begin{displaymath}}
\newcommand{\een}{\end{displaymath}}
\newcommand{\beqn}{\begin{equation}}
\newcommand{\eeqn}{\end{equation}}
\newcommand{\beqna}{\begin{eqnarray*}}
\newcommand{\eeqna}{\end{eqnarray*}}

\newcommand{\nn}{\nonumber} 

\def\supp{\operatorname{supp}}

\newcommand{\sfrac}[2]{\textrm{\footnotesize $\frac{#1}{#2}$}}

\newcommand{\inn}[1]{\langle {#1} \rangle }

\newtheorem{lemma}{Lemma}
\newtheorem{theorem}[lemma]{Theorem}
\newtheorem{remark}[lemma]{Remark}
\newtheorem{proposition}[lemma]{Proposition}
\newtheorem{corollary}[lemma]{Corollary}
\newtheorem{definition}[lemma]{Definition}
\newtheorem{hyp}{Hypothesis}

\numberwithin{equation}{section}
\numberwithin{lemma}{section}

\makeatletter
\newsavebox{\@brx}
\newcommand{\llangle}[1][]{\savebox{\@brx}{\(\m@th{#1\langle}\)}%
  \mathopen{\copy\@brx\kern-0.5\wd\@brx\usebox{\@brx}}}
\newcommand{\rrangle}[1][]{\savebox{\@brx}{\(\m@th{#1\rangle}\)}%
  \mathclose{\copy\@brx\kern-0.5\wd\@brx\usebox{\@brx}}}
\makeatother

\newlength{\hght}
\newlength{\dpth}

\begin{document}
\title{ On asymptotic  expansions   of resolvents   for  Poisson distributed random Schr\"odinger operators}
\author{\vspace{5pt} David Hasler$^1$\footnote{
E-mail: david.hasler@uni-jena.de } \quad  Jannis Koberstein$^1$\footnote{E-mail: jannis.koberstein@eah-jena.de} \\
\vspace{-4pt} \small{$1.$ Department of Mathematics,
Friedrich Schiller  University  Jena} \\ \small{Jena, Germany } }
\date{}
\maketitle

\begin{abstract}
We study expectation values of  matrix elements  of the resolvent   for a random   Schr\"odinger operator with  potential distributed according 
to a Poisson process. Asymptotic expansions for  these matrix elements of resolvents  in the limit of small disorder are derived.  Explicit  
estimates for the expansion coefficients are given and  we show that their infinite volume limits  
are   finite as   the spectral parameter approaches the spectrum 
of the free Laplacian.   
\end{abstract}

\section{Introduction}

Solids occur in nature in various forms and are almost always subject to disorder of some form. Sometimes they are (almost) totally ordered, sometimes they are (more or less) completely disordered.  A random Schr\"odinger operator is  a  
mathematical caricature  to describe   quantum mechanical aspects of such random systems in an approximation where many-body
interactions  are neglected. Anderson introduced a random Schr\"odinger operator where the potential was given by a sum of potentials 
arranged in a lattice with random strength at each lattice site \cite{Anderson.1958}.  
 In this paper, we study a random Schr\"odinger operator where the  randomness is given by a 
Poisson distributed random potential with random weights.  Such  random potentials are used to  model amorphous materials like glass or rubber.   

An analysis of matrix elements of  resolvents of Schr\"odinger operators is 
 important,  since they   charactarize spectral properties such as    spectral measures.
Spectral measures of Schrödinger operators encode important spectral properties. 
 They can be  used to determine  spectral types, such as point 
 spectrum  or absolutely continuous spectrum, which are closely linked 
 to physical phenomena like bounded  or extend states. 
 These states are key to understanding the dynamics of particles and are
  therefore essential  in the study of properties like electrical conductance.
In addition, the integrated density of states (IDS), a key quantity, 
 can be expressed as the expectation of a normalized sum of spectral measures. 
For random Schrödinger operators
the IDS plays a central role, as it provides the average number of eigenstates (per unit volume) below a given energy level. 
 On the other hand, expectations of powers  of spectral measures can be used 
 to detect  extended states.   In particular, in   the small disorder regime, this turns out
 to be   a challenging 
subject, and has beend addressed in several publications, cf. 
\cite{Loyd.1969,AcostaKlein.1992,MR1444083, Klein.1998,Schenker.2004,AizenmanSimsWarzel.2006,FroeseHaslerSpitzer.2007,
ErdosSalmhoferYau.2007b,ErdosSalmhoferYau.2008a,AizenmanWarzel.2013,KirschKrishna.2020,AnantharamIngremeauSabriWinn.2021} and references therein.

In the present paper,  we consider expectations of matrix elements of the resolvent of a random Schr\"o\-dinger operator with a random potential  given 
by a Poisson measure. In particular,  we are interested in this quantity for small  disorder and its behavior as the disorder strength tends to 
zero. It is expected, that in this limit, one recovers the corresponding quantity for the free Schr\"odinger operator. We show 
that on a  certain energy  scale this is indeed the case.  More explicitly, we show  that the expectations have an asymptotic expansion as the disorder strength tends to zero in a certain  scaling limit. The expansion
is obtained by  means of a Neumann series.
We note that a related  result  has been obtained recently by the present authors  \cite{HaslerKoberstein.2024}.
In   \cite{HaslerKoberstein.2024} the resolvent was approximated for small but nonzero disorder  in terms of smeared out expressions,
and the error estimate
was obtained by means of 
 a Duhamel expansion.
 Furthermore,  we show in the present paper  that the 
infinite volume limits of the expansion coefficients are finite as the spectral parameter approaches the real axis, inside the spectrum
of the free Laplacian.  For this, we derive  formulas expressing the expansion coefficients as 
 convergent integrals. 
 We then combine these two main results to obtain a concrete asymptotic expansion in the infinite volume limit. 
 For this we make use of the existence of the infinite volume limit shown in \cite{HaslerKoberstein.2024-1}.
  In a following paper we will prove an  asymptotic expansion for    the integrated density of states, based 
 on estimates given in the present paper. In that sense the present paper serves as the foundation or a first part for the mentioned follow-up work.

We note that a related expansion of the resolvent has been studied in  
\cite{MagnenPoirotRivasseau.1996,MagnenPoirotRivasseau.1998,Poirot.1999}.
In these works the asymptotic for small disorder for a random Schr\"odinger operator in two  dimensions  with a random potential given by a Gaussian random field was studied.  Moreover, we want to mention deterministic perturbation results  about  distributions of eigenvalues for periodic Schr\"odinger operators \cite{FeldmannKnoerrerTrubowitz.1990,FeldmannKnoerrerTrubowitz.1991}. The methods we use in the present paper are inspired by 
the expansion techniques used in \cite{ErdosSalmhoferYau.2007a,ErdosSalmhoferYau.2007b,ErdosSalmhoferYau.2008a}.  
Our results on the expansion coefficients and  their  infinite volume limit 
are based on  the method of analytic dilation,  cf.  \cite{CombesThomas.1973,SR4} and references therein. 
Furthermore, we want to point out the result \cite{MR1444083} on the density of states for a Schr\"odinger operator  with a Poisson distributed random potential, 
where an  asymptotic expansion  in the  concentration strength was studied in  distributional sense.

The structure of the present  paper is as follows. In Section \ref{Model}  we introduce the model and state the main results 
about expectations of matrix elements of resolvents. 
In Section \ref{x_asymptotic} we use the main  results to derive a concrete asymptotic expansion. 
In Section \ref{PPL} we  study  expectations of products of the random potential, which will be needed for the proofs.
In Section  \ref{sec:prooasymexp} we prove results about the expansion coefficients for expectations of  matrix elements of the resolvent and show that for their infinite 
volume limit the spectral parameter can be extended continuously to the real axis. 
In Section \ref{sec:AsymEE} we then 
show that  these  expansion coefficients indeed yield an  asymptotic expansions of expectations of matrix values of the resovlent. 
Finally the appendix consists of three parts.  Part \ref{AppendixA} uses a resolvent identity to derive a Neumann expansion for the inverse of the sum of two operators. Part \ref{lemmanadilpotproof} gives examples of dilatation analytic functions. Finally Part \ref{AppendixC} states estimates on integrals an gives a way to estimate sums by integrals.

\section{Model and main  Results} 
\label{Model}
In this section we  introduce  the model and state the main results about the expectation 
of the resolvent.  We note that the definition of the model follows  the one given in \cite{ErdosSalmhoferYau.2008a} closely.
We consider   a finite box of size $L$,  with $L  > 0$, and define the  box 

 $\Lambda_L := \left[ - \frac{1}{2} L , \frac{1}{2} L \right)^d \subset \R^d$. 
Let $\inn{ \cdot , \cdot }_{L^2(\Lambda_L)}$ and $\| \cdot \|_{L^2(\Lambda_L)}$ denote the canonical scalar product and the norm of the Hilber space  $L^2(\Lambda_L)$. If it is clear from the 
context what the inner product or  the norm is,   we shall occasionally drop the subscript $L^2(\Lambda_L)$.
 The kinetic energy will be given by the periodic Laplacian. To introduce this operator it is convenient to 
work in terms of Fourier series. Let  
$\Lambda_L^*  = ( \frac{1}{L} \Z  )^d=\{ (k_1 ,  \ldots ,  k_d) : \forall j=1, \ldots , d,  \exists m_j \in \Z,: k_j =\frac{m_j}{L} \}$ denote  
the so-called dual lattice. We introduce the notation 
\begin{equation}
\int_{\Lambda_L^*} f(p) dp  := \frac{1}{|\Lambda_L |} \sum_{p \in \Lambda_L^*} f(p) . 
\end{equation} 
The sum  $\int_{\Lambda_L^*}f(p)dp$ can be interpreted as a Riemann-sum, which  converges to the integral $\int f(p) dp $ 
as   $L \to \infty$, provided $f$ has  sufficient  decay at infinity and sufficient regularity.  
For any $f \in L^1(\Lambda_L)$ we define the Fourier series 
$$
\hat{f}(p) = \int_{\Lambda_L} e^{ - 2 \pi i p \cdot x}  f(x) dx \text{ for } p \in \Lambda_L^*  . 
$$
  For $g    \in \ell^1(\Lambda_L^*)$ we define 
$$
\check{g}(x) = \int_{\Lambda^*_L} e^{  2 \pi i  p \cdot x } g(p) dp  = \frac{1}{|\Lambda_L |} \sum_{p \in \Lambda_L^*} e^{  2 \pi i  p \cdot x } g(p) \text{ for }  x \in \Lambda_L. 
$$
Note that 
$\check{(\cdot)}$ extends uniquely to a continuous linear map  on $ \ell^2(\Lambda_L^*)$.  
We shall denote this extension again by the same symbol. This extension maps $\ell^2(\Lambda_L^*)$ unitarily to $L^2(\Lambda_L)$
and is the inverse of $\hat{(\cdot)}|_{L^2(\Lambda_L)}$, see for example \cite[Theorem 8.20]{folland}.  From a physics background one might be familiar with the notation where  $L^2(\Lambda_L)$ represents the position space,  while $\ell^2(\Lambda_L^*)$ would be called momentum space,  or Fourier space.
For $ p \in \Lambda_L^*$  and $x \in \Lambda_L$ we define 
\begin{align} \label{eq:deofonb}
 \varphi_p(x) ={|  \Lambda_L|}^{-1/2} e^{ i 2 \pi p \cdot x } .
\end{align}
Observe that 
$
\{ \varphi_p :p \in \Lambda_L^* \} 
$
is an orthonormal Basis (ONB) of $L^2(\Lambda_L)$ \cite[Theorem 8.20]{folland} and that 
$\hat{f}(p) = | \Lambda_L|^{1/2}  \inn{  \varphi_p , f}_{L^2(\Lambda_L)} $.

When defining the Laplacian restricted to a finite box,  we have to choose boundary condition to describe how our operator behaves at the border of the box.  
We are presented with the choice between 
Dirichlet, Neumann and periodic boundary conditions.
Note that in  the  sense of forms,  periodic boundary conditions lie  between Dirichlet and Neumann boundary conditions,  where the latter two are better suited,  when working in position space.  
While we will be starting in position space, the majority of this work
will be performed in momentum space.,  where  using periodic boundary conditions is technically convenient.  
Therefore,  we are going introduce the Laplacian with periodic boundary conditions on $L^2(\Lambda_L)$ by means of Fourier series.   
In this paper we shall adapt the physics convention that for a vector $a =(a_1,...,a_d)$ we use the notation $a^2 := |a|^2$ where $|a| :=  \sqrt{ \sum_{j=1}^d |a_j|^2}$.  
Using this notation,  we define the energy function
\begin{equation}
\label{energiyfunction}
\nu : \R^d  \to \R_>,  \quad p \mapsto \nu(p) := \frac{1}{2} p^2,   
\end{equation} 
which we can now use to
 define $-\Delta_L$ as the linear operator with domain 
\begin{align*}
D(-\Delta_L) = \{ f \in L^2(\Lambda_L) : \nu \hat{f} \in \ell^2(\Lambda_L^*) \},
\end{align*} 
and the mapping rule
\begin{equation} \label{eq:deoflaplac0} 
 -\Delta_L : D(-\Delta_L)  \to L^2(\Lambda_L), \quad f \mapsto -\Delta_L f   =  (2 \pi )^2 ( \nu  \hat{f})^\vee . 
\end{equation} 
Observe that  $-\Delta_L$ is selfadjoint, since it is  unitary equivalent to  a multiplication operator by a real valued function,  and that we have the identity 
\begin{equation} \label{eq:deoflaplac} 
\left[ - \frac{1}{2 (2 \pi)^2}   \Delta_L  f \right]^\wedge(p)  = \nu (p) \hat{f}(p)  .
\end{equation} 

By  
\begin{equation} \label{defofmainH}  
 H_{L} := H_{\lambda,L} := - \frac{\hbar^2}{ 2 m } \Delta_L + \lambda V_{L}
\end{equation} 
we denote a random Schr\"odinger operator acting in $L^2(\Lambda_L)$ with   a random potential
$
V_{L} = V_{L,\omega}(x)
$, defined below,  and a  coupling constant $\lambda \geq 0$.   As in  \cite{ErdosSalmhoferYau.2008a}  we choose units for the mass $m$ and Planck's constant  so that $\frac{\hbar^2}{2m} = [2 (2\pi)^2]^{-1}$.
For a function $h : \R^d \to \C$ we denote by $h_\#$ the $L$--periodic extension of $h|_{\Lambda_L}$ to $\R^d$. 
The potential is given by
\begin{align} \label{defofpot101} 
V_{L,\omega}(x) := \int_{\Lambda_L} B_\#(x - y) d \mu_{L,\omega}(y),
\end{align} 
where $B$,  having the physical interpretation as   a single site potential profile,   is  assumed  to be a  real valued Schwartz function on $\R^d$.
Moreover, we assume that either $B$  has   compact support
or that  $B$  is  symmetric with respect to the reflections of the coordinate axis,  i.e.  for  $j=1,...,d$
\muun{ \label{SC}
\mathfrak{S}_j:  (x_1,...,x_j,...,x_d) &\mapsto (x_1,...,-x_j,...,x_d) 
\\ 
B \circ \mathfrak{S}_j &= B. \nn
}  
Note that  we  do not assume that $B$ is nonnegative, as such a property will not be  
 required  for our analysis.

\begin{remark}{\rm 
Each of the two conditions is mathematically convenient in the sense that they ensure sufficiently fast decay of the Fourier transform.  While the reflection symmetry condition is satisfied by rationally invariant potentials, which occur naturally. }
\end{remark}
Furthermore,  $\mu_{L,\omega}$ is a Poisson point measure  on $\Lambda_L$ with homogeneous unit density and 
with independent indentically distributed (i.i.d.) random masses.  
More precisely, for almost all events $\omega$, it consists of $M$ points $\{ y_{L,\gamma}(\omega) \in \Lambda_L : \gamma=1,2,...,M\}$, where $M = M(\omega)$ is a Poisson variable with expectation $|\Lambda_L|$, and $\{ y_{L,\gamma}(\omega) \}$ are i.i.d. random variables uniformly distributed on $\Lambda_L$.   Both are independent of  the random i.i.d. weights  $\{ v_\gamma : \gamma=1,2,... \}$ and 
the random measure is given by 
$$
\mu_{L,\omega} = \sum_{\gamma=1}^{M(\omega)} v_\gamma(\omega) \delta_{y_{L,\gamma}(\omega)} , 
$$ 
where $\delta_y$ denotes the Dirac mass at  the point $y$.
Note that the case where $M(\omega )=0$ corresponds to a vanishing potential.
Furthermore we denote the the common  distribution of the  weights   $\{ v_\gamma \}$  by  ${\bf P}_v$ and assume that the moments 
\muun {\label{moments}
m_k :=  {\bf E} v_\gamma^k
}
 satisfy
\begin{equation} \label{2.4} 
 m_{k } < \infty     , \quad  \text{ for all }  k \in \N.
\end{equation} 
For some of the results we will use that the first moment $m_1 = {\bf E} v_\gamma$ vanishes.
Note that for convenience the notation  will not reflect the   dependence on the specific  choice of the random distribution ${\bf P}_v$. 
Observe that we can write   \eqref{defofpot101} as  
\begin{equation}\label{bumpnotationres} 
 V_{L,\omega}(x) =  \sum_{\gamma=1}^M V_{L,\gamma}(x)   \quad \text{ with } \quad V_{L,\gamma}(x) := v_\gamma B_\#(x-y_{L,\gamma}) .
\end{equation} 
The expectation with respect to the joint measure of $\{ M , y_{L,\gamma}, v_\gamma\}$ is denoted by ${\bf E}_L$.
Sometimes we will use the notation  
\begin{equation}\label{esy:3.24} 
{\bf E}_L = {\bf E}_M {\bf E}_{y_L}^{\otimes M } {\bf E}_v^{\otimes M } 
\end{equation} 
referring to the expectation of $M$, $\{ y_{L,\gamma} \}$ and $\{ v_\gamma \}$ separately. 
In particular, ${\bf E}_y^{\otimes M }$ stands for the normalized integral 
\begin{equation}\label{esy:3.240} 
\frac{1}{|\Lambda_L|^M} \int_{(\Lambda_L)^M} dy_1 \cdots dy_M . 
 \end{equation} 

Since the potential is almost surely bounded it follows  by standard perturbation theorems, e.g. the Kato-Rellich theorem \cite{SR2},  that   the operator \eqref{defofmainH}  is almost 
surely self-adjoint  for all $\lambda \geq 0$.

The first object of our interest is the  expectation  of  matrix values of the resolvent
\muun{
\label{GFct}
{\bf E}_L  \langle \psi_1, ( H_{ L}  - z)^{-1}   \psi_2 \rangle \quad \text{ for } \psi_1 , \psi_2 \in L^2 (\Lambda_L)
}
as the spectral parameter approaches the real axis. 
Note that  \eqref{GFct}  is the expectation of the interacting resolvent, which is difficult to study.  To analyze  \eqref{GFct}  we will  use a Neumann-Type expansion to express the interacting  resolvent as the sum of products of powers of the potential and the resolvent of the periodic Laplacian, $\Delta_L$. This  is the content of   Lemma \ref{eq:resexp} below. 
For notational compactness   we  shall denote   the resolvent of $-\Delta_L$  by 
$$
R_L(z ) :=  \left(- \frac{\hbar^2 }{2 m} \Delta_L - z \right)^{-1} , 
$$
where $z \in \C \setminus [0,\infty)$.
Note that    we use a   notation for the   resolvent of the free Laplacian, which   one might expect  for the interacting  one.  
Iterating the second resolvent identity \ref{ResId}  yields the following lemma.

\begin{lemma} 
For $z \in \C \setminus [0,\infty)$, $L > 0$, and $n \in \N$ we have
\begin{align} \label{eq:resexp} 
( H_L  - z)^{-1} & = \sum_{j=0}^n R_L(z) [ \lambda V_L R_L(z)  ]^j +  [R_L(z)  \lambda V_L  ]^{n+1}  ( H_L  - z)^{-1} . 
\end{align}
\end{lemma}
\begin{proof}
This follows directly from Lemma \ref{resolventenformel_interiert}  with $A=- \frac{\hbar}{2m}\Delta_L - z$ and $B=\lambda V_L$, since  $\sigma (H_L) \subset [0,\infty)$ see for example  \cite{RandOp}.
\end{proof}
With \eqref{eq:resexp} on our hands, we can use the linearity of both the scalar product in each entry  as well as the expectation  value.  With this we can now study the following expressions.
 For  $z \in \C \setminus [0,\infty)$ and $\psi_1, \psi_2 \in L^2(\R^d)$ we define
 \begin{align} \label{defoffinT} 
  T_{n,L}[z;\psi_1,\psi_2] 
& :=  {\bf E}_L \inn{ \psi_{1,\#},    R_L(z) [  V_L  R_L(z)  ]^n \psi_{2,\#} }  . 
\end{align}
By definition the potential is almost surely bounded. Thus  \eqref{defoffinT} is well defined almost surely.
We are now going to work towards Theorem \ref{lem:formulaforexpcoeff}, which will express   \eqref{defoffinT}  in momentum space.
For this, we  introduce  the discrete delta function  in momentum space  for $u \in \Lambda_L^*$
\begin{equation}
\delta_*(u) = 1_{\{0\}}(u) 
\end{equation} 
and a  normalized discrete delta function 
\begin{equation} \label{deltaL}  
\delta_{*,L}( u  ) = |\Lambda_L|  \delta_*(u),
\end{equation} 
where we used  the notation that for a set $A$ we write $1_A(x) = 1$, if $x \in A$, and $1_A(x) = 0$, if $x \notin A$.  \\

We now need to consider  the expectation of the product of the   random variable  $V_{L}$ i.e.
$
 \textbf{E}_L \left[ \prod_{j=1}^n v_{\gamma_j}\right]
$,  for they will appear from \eqref{defoffinT}.
This topic will be discussed in more detail in Section \ref{PPL}. The short idea is that we can use independence of the variables as long as they are different, while for the ones that are the same,  we count how often each of the variables occurs: Multiple appearances of the same random variable will result in a higher moment of the respective variable after evaluating the expected value.

To express the result of this procedure  we introduce the following partition function.  It (in a sense) filters all possible partition and leave only one remaining,  which represents the configuration of the momenta.  Meaning that every set $a$ in the partition $A$ has to correspond  to exactly one site in the sense that for  all $ a\in A $ there is a unique $y_a \in \Lambda_L$ such that $y_\gamma = y_a$ for all $\gamma \in a$.
To do so let $\mathcal{A}_n$,  denote the set of partitions of the set $\{1,...,n\}$.  
For $A \in \mathcal{A}_n$ and $k= (k_1,...,k_{n+1})$ we define the partition function
\begin{align} \label{eq:defofdeltapi0}
\mathcal{P}_{A,L} : (\Lambda_L^*)^{n+1} \to \R,  \ k  \mapsto
\mathcal{P}_{A,L}(k)  =  \prod_{a \in A} \left\{ m_{|a|}  \delta_{*, L} \left( \sum_{l \in a} (k_l - k_{l+1}) \right) 
  \prod_{l \in a}  \widehat{B}_\#(k_l - k_{l+1} ) \right\}.
 \end{align}
Recall  that  $m_{|a|}$ is the $|a|$-th moment defined in \eqref{moments}.
To state the results we  will  define the infinite volume Fourier transform and inverse Fourier transform for $f \in L^1 (\R^d) $
\begin{align*}
\hat{f}(k) &= \int_{\R^d} f(x) e^{ - 2 \pi i k \cdot  x } dx  , \quad k \in \R^d , \\
\check{f}(x) &=\int_{\R^d}   f(k) e^{  2 \pi i k \cdot  x } dk, \quad x \in \R^d , 
\end{align*}
repectively.

For the next theorem we transition over to the momentum space, where we  give an explicit expression for the  expansion coefficients.  Remember that we write $k= (k_1, \ldots k_{n+1}) \in (\Lambda_L^*)^{n+1}$ and that the real energy function $\nu$ defined in \eqref{energiyfunction} is non-negative.

\begin{theorem} \label{lem:formulaforexpcoeff} 
For all $z \in \C \setminus [0,\infty)$ and $\psi_1 ,\psi_2 \in L^2(\R^d)$ we have 
\begin{align*}
  T_{n,L}[z;\psi_1,\psi_2] 
& =   \int_{(\Lambda_L^*)^{n+1}} \sum_{A  \in \mathcal{A}_n} \mathcal{P}_{A,L}((k_1, \ldots , k_{n+1}))   \overline{\widehat{\psi}}_{1,\#}(k_1)    
 \widehat{\psi}_{2,\#}(k_{n+1})
 \\
 &\times \prod_{j=1}^{n+1}  (\nu( k_j) -  z)^{-1} d (k_1, \ldots , k_{n+1}).
\end{align*}
\end{theorem} 

The proof of Theorem \ref{lem:formulaforexpcoeff}  will be  given in Section \ref{sec:prooasymexp}. The next theorem shows that 
for each expansion coefficient  the infinite volume 
limit   exists and  it gives for each expansion coefficient an upper bound dependent on  the spectral parameter.

\begin{theorem}  \label{exinflim} Let  $n \in \N$. 
For all  $z \in \C \setminus [0,\infty)$, and   $\psi_1 ,\psi_2 \in L^2(\R^d)$  the   limit 
$ \lim_{L \to \infty} T_{n,L}[z; \psi_1, \psi_2] $
exists in $\C$. There exists a  constant $K_n$ such that  the inequality 
\begin{align} \label{boundonTs} 
| T_{n,L}[z;\psi_1,\psi_2 ] | \leq \frac{ K_n   \| \psi_1 \| \| \psi_2 \| }{ {\rm dist}(z,[0,\infty))^{n+1} }    
\end{align} 
holds for all $L \geq 1$, $z \in \C \setminus [0,\infty)$, and  $\psi_1, \psi_2 \in L^2(\R^d)$.
\end{theorem} 
Theorem    \ref{exinflim} will follow as an immediate consequence 
of Proposition  \ref{propexofcinf},  shown  in Section \ref{sec:prooasymexp} as well.
By  the first statement  of  Theorem    \ref{exinflim} we can define the infinite volume 
 limit of the expansion coefficients  
\begin{equation} \label{eq:defoflimL} 
T_{n,\infty}[z ; \psi_1,\psi_2] := \lim_{L \to \infty} T_{n,L}[z; \psi_1, \psi_2]  ,
\end{equation}
for any $\psi_1,\psi_2 \in L^2(\R^d)$  and  $z \in \C \setminus [0,\infty)$. 
To state the   main result  about the 
expansion coefficients of the resolvent  \eqref{eq:defoflimL},      
  we need some regularity assumptions for the profile function and the wave functions with respect to which 
  we calculate the matrix element of the resolvent. These regularity assumptions are   
  formulated in terms of so called analytic dilations, cf. \cite{CombesThomas.1973,SR4}.  Let us first introduce the group of dilations.
 
 \begin{definition}
 \label{defutheta}
The group of unitary operators $u(\theta)$ on $L^2(\R^d)$ given by 
\muun{
\label{equtheta}
(u(\theta) \psi)(x) = e^{ \frac{d \theta}{2} } \psi(e^\theta x ) \quad \text{for} \ \theta \in \R , 
}
 is called the {\bf group of dilation operators on} $\R^d$. 
 \end{definition}
 
 It is straight forward to verify that $\R \mapsto \mathcal{B}(L^2(\R^d))$, $\theta \mapsto u(\theta)$ is indeed a strongly continuous one paramater group of  unitary operators, cf. \cite[Page 265]{SR1}. 
 Observe that from the definition of the Fourier transform and the dilation operator it is easy to see that  for any $\psi \in L^2(\R^d)$ and all $\theta \in \R$ 
 \begin{equation} \label{eq:dilfour}  
 (u(\theta) \psi)^{\wedge} = u(-\theta) \hat{\psi}. 
 \end{equation} 
Let us first state the hypothesis about the profile function needed for Theorem  \ref{thm:boundaryvelueexpcoeff}, below.

\begin{hyp} \label{H1} There exists $\vartheta_B > 0$ such that the Fourier transform, $\hat{B}$, of the Schwartz function $B$ satisfies that for $p=1 $ and $p=\infty$ the  function $\R \to L^p(\R^d),  \theta \mapsto  \widehat{B}_{\theta} := (u(\theta) B)^\wedge$  has  an extension 
to an analytic function
$D_{\vartheta_B} := \{ z \in \C : |z| < \vartheta_B\}\to L^p(\R^d).$
\end{hyp} 

Let us now state the hypothesis about the wave function with respect to which we shall calculate matrix elements of the resolvent  in Theorem \ref{thm:boundaryvelueexpcoeff}  below.

\begin{hyp} \label{H2} For the vector $\psi \in L^2(\R^d)$ and there exists $\vartheta_\psi > 0$ such that the transform $\hat{\psi}$ satisfies that the  function  $\R \to L^2(\R^d)$,    $\theta \mapsto \widehat{\psi}_\theta := \widehat{\psi}(e^\theta \cdot ) $ has  an extension 
to an analytic function $D_{\vartheta_\psi}  := \{ z \in \C : |z| < \vartheta_\psi\} \to L^2(\R^d)$.
\end{hyp}

In Appendix \ref{lemmanadilpotproof}  Lemma \ref{lem:anaprop} we gives a  class of functions 
  satisfying   these  hypotheses.  For example if $B$ or $\psi$, respectively, is a product  of  a polynomial, a  Gaussian, and   a free wave function with any wave vector then it satisfies  Hypotheses \ref{H1} or \ref{H2}, respectively.

We can now state the first main result, which establishes the existence of the boundary value of
 the expansion coefficients. 
We shall use the following notations
\begin{align*}
\R_> := \{ x \in \R : x > 0 \} , \quad \HS := \{ z \in \C : {\rm Im} (z) > 0 \}.
\end{align*} 

\begin{theorem} \label{thm:boundaryvelueexpcoeff} Suppose  Hypothesis \ref{H1} holds and suppose $\psi_1,\psi_2 \in L^2(\R^d)$ satisfy  Hypothesis   \ref{H2}. Then for any $E > 0$ the following limit exists    as a finite complex number 
$$
 T_{n,\infty}[E \pm  i 0^+;\psi_1,\psi_2] := \lim_{\eta \downarrow 0 } T_{n,\infty}[E \pm  i \eta;\psi_1,\psi_2] .
$$
Moreover,  $z \mapsto T_{n,\infty}[z  ; \psi_1, \psi_2] $  as a function on $\HS$  has a continuous extension to a function on $ \HS \cup  \{ z \in \C : { \rm Re} (z) > 0 \text{ and } {\rm Im} (z) = 0 \}$.
\end{theorem} 
The proof of  Theorem \ref{thm:boundaryvelueexpcoeff}  will be given in Section     \ref{sec:prooasymexp}.  
The next theorem is  the second main result about  expectations of matrix elements  of the resolvent.   It shows  
that the expectations of matrix elements of the resolvent have an asymptotic expansion with expansion coefficients given by \eqref{defoffinT}. The estimate is uniform 
in the size  $L \geq 1$  of the box and yields an asymptotic expansion  as the spectral parameter approaches the positive real axis  in a the weak coupling limit where $\eta= \lambda^{2-\epsilon}$.

\begin{theorem} \label{thm:maintec000}  Assume  $m_1 = \E v_\gamma = 0$.  Then there exists an $L_0 \geq 1$ with the following property. 
\begin{itemize}
\item[(a)] For all  $n \in \N$ and $E > 0$ there exists a constant $K_{n,d,E,B} $  such that for all $\psi_1,\psi_2  \in L^2(\R^d)$, $\eta > 0$, and $L \geq L_0$  we have  
\begin{align}
& \left|  \E \inn{ \psi_{1,\#} , (H_{  \lambda ,  L} - E  \mp  i  \eta  )^{-1}      \psi_{2,\#} }_{L^2(\Lambda_L)}  -  \sum_{j=0}^{n-1} T_{j,L}[E  \pm    i \eta  ; \psi_1, \psi_2]   \lambda^j  \right| \nonumber \\
&  \leq K_{n,d,E,B}  \left( \frac{ \lambda^2  }{\eta}   \right)^{n/2} \left(1 +   \ln(   \eta^{-1} + 1 )\right)^n    \eta^{-{3/2}}  \| \psi_1 \| \| \psi_2 \|. \label{eq:boundonexp}  
\end{align} 
Furthermore, the constant  $K_{n,d,E,B} $ as function of $E$ can be chosen to be uniformly bounded  on compact subsets of $\R_>$.
\item[(b)] 
For  all  $\epsilon \in ( 0,2)$,   $N  \in \N$  and compact subset $I \subset \R_>$  there 
exists a constant $C$ such that  for all $\psi_1,\psi_2  \in L^2(\R^d)$,  $\lambda \geq 0$, $L \geq L_0$, and $E \in I$ 
\begin{align*} 
& \left|  \E \inn{ \psi_{1,\#} , (H_{  \lambda ,  L} - E \mp   i \lambda^{2 - \epsilon} )^{-1}      \psi_{2,\#} }_{L^2(\Lambda_L)}  -   
 \sum_{n=0}^{ \lceil 4 (N+3)/\epsilon \rceil} T_{n,L}[E \pm    i \lambda^{2 - \epsilon} ; \psi_1, \psi_2]    \lambda^n \right|  \\
 & \leq   C  \lambda^N   \| \psi_1 \| \| \psi_2 \| .  
\end{align*} 
\end{itemize} 
\end{theorem}
The proof of  Theorem \ref{thm:maintec000}  will be given in Section     \ref{sec:AsymEE}.
We note that the $T_{0,L}$-term is simply given by the free Laplacian, with periodic 
boundary conditions, which in the limit as $L$ tends to infinity tends to the Laplacian on $\R^d$. 

\begin{remark} {\rm  The energy  $E > 0$ lies inside the spectrum of the free Laplacian which  presents the challenge we want to address in this paper. 
Note that for negative energies $E < 0$ outside of the free Laplacian, the corresponding  estimates and limits of Theorems \ref{thm:boundaryvelueexpcoeff} and \ref{thm:maintec000}  are in fact 
straight forward  to establish   by means of the bound   \eqref{boundonTs}}.
\end{remark} 

\begin{remark} {\rm Regarding the choice of $L_0$ in Theorem \ref{thm:maintec000} we note that in the case where the profile function $B$   satisfies the symmetry condition \ref{SC}  Theorem \ref{thm:maintec000} holds for  $L_0 =1$. In case  $B$ has compact support,  one chooses $L_0\geq 1$ such that $\supp B \subset (-L_0/2,L_0/2)^d$.}
\end{remark}

\begin{remark} {\rm An estimate similar to  \eqref{eq:boundonexp}   was obtained in  \cite{MagnenPoirotRivasseau.1998} for a related model. In  that work
a Laplacian with  a Gaussian random potential in two dimensions was studied and it was shown that the difference between the expectation of the  resolvent  and a renormalized free resolvent  is  operator norm bounded by   $\left( \sfrac{\lambda^{2 + \delta}}{\eta} \right)^3 \frac{1}{\eta}$ for some 
$\delta > 0$,  provided $0 \leq \eta \leq \lambda^2$.   Now the bound   \eqref{eq:boundonexp}  in Theorem \ref{thm:maintec000} can be made 
much smaller than $\left( \sfrac{\lambda^{2 + \delta}}{\eta} \right)^3 \frac{1}{\eta}$ in  the regime where $\eta = \lambda^{2 - \epsilon}$  by  choosing  $n$ sufficiently large. }
\end{remark} 

\begin{remark} {\rm  It would be interesting to investigate to what extent the  expansion  \eqref{eq:boundonexp}   in Theorem \ref{thm:maintec000} could be improved 
using  so called tadpole renormalization and an analysis of crossing graphs,  see for example \cite{ErdosSalmhoferYau.2008a}.  In that paper this was successfully  used to establish quantum diffusion in a diffusive  scaling limit for similar model to the one introduced in the present paper.  We did not perform such an analysis in this work,  
since the focus of it is on the expansion coefficients to  any order. 
}
\end{remark}

\begin{remark}{\rm  We note that in 
\cite{HaslerKoberstein.2024}  a   result has been obtained  similar to Theorem \ref{thm:maintec000}.
In \cite{HaslerKoberstein.2024}  an approximation of matrix elements 
in terms of smeared out coefficients  was shown 
to arbitrary precision for small but fixed $\lambda$. To show the 
results in  \cite{HaslerKoberstein.2024}  
 a DuHamel expansion was used. 
 In the present paper 
we prove the existence of 
an asymptotic expansion for $\lambda \downarrow 0$ and use a Neumann series. 
}
\end{remark}

\section{Asymptotic Expansion}

\label{x_asymptotic} 

In this section we combine the  results of the main Theorems \ref{thm:boundaryvelueexpcoeff} and  \ref{thm:maintec000}  
 to   derive a  concrete  asymptotic expansion   for  infinite volume limits of  matrix elements of resolvents. 
We note that  the assumption made in  \eqref{defoflimit-00} of the following Theorem  has been verified in 
\cite{HaslerKoberstein.2024-1}.

\begin{theorem}\label{asymptotictheorem} Assume Hypothesis \ref{H1},  $\E v_\gamma = 0$, and  $\psi_1,\psi_2 \in L^2(\R^d)$ 
satisfy Hypothesis   \ref{H2}.   Let  $E > 0$,  $\lambda_0 > 0$,  and assume that the limit 
\begin{align} \label{defoflimit-00}
R_{\psi_1,\psi_2}(\lambda;  E\pm i \eta) := \lim_{L \to \infty} \E \inn{ \psi_{1,\#} , (H_{  \lambda ,  L} - E  \mp  i  \eta  )^{-1}      \psi_{2,\#} }_{L^2(\Lambda_L)} 
\end{align} 
exists for all $\lambda \in [0,\lambda_0]$.  
Let $\epsilon \in (0,2)$  and $N \in \N$. Then there exists 
a constant $C$ such that with $\eta = \lambda^{2 - \epsilon}$   for all  $\lambda \in [0, \lambda_0]$ 
\begin{align} \label{expansionasymptoticmain}
 &  \Bigg| R_{\psi_1,\psi_2}(\lambda;E\pm i \eta)  -     \sum_{n=0}^{N-1}   T_{n,\infty}[  E \pm i \eta ; \psi_1, \psi_2] \lambda^n
\Bigg|   \leq  C \lambda^{N}  .
\end{align} 
\end{theorem}

\begin{proof} 
In view of the bound in  \eqref{boundonTs}  and the bound   in \eqref{ln-1} together with  the spectral theorem,
it suffices to assume  $\lambda_0 \leq 1$. 
Now  using Theorem   \ref{thm:maintec000} (b) and then taking  subsequently the limit  
 $L \to \infty$  by  means of \eqref{defoflimit-00} and  Theorem   \ref{exinflim}
we see that  
 \begin{align}  \label{bounddensity-0} 
 & \left|  R_{\psi_1,\psi_2}(\lambda;E\pm i \eta)   - 
  \sum_{n=0}^{ \lceil 4 (N+3)/\epsilon \rceil}  T_{n,\infty}[E \pm    i \eta ; \psi_1 , \psi_2]   \lambda^n 
 \right| \leq   C_0 \lambda^{N} 
\end{align} 
for some constant $C_0$ and $0 \leq \lambda \leq 1$. 
By means of  Theorem \ref{thm:boundaryvelueexpcoeff}, we know that
\begin{align}  \label{regofotherTs-2-0}
C_{(n)} := \sup_{\eta \in [-1,1] \setminus\{0 \} }  |T_{n,\infty}[E +    i \eta ; \psi_1, \psi_2] | < \infty  . 
\end{align} 
Now      using  the triangle inequality, \eqref{regofotherTs-2-0},  and \eqref{bounddensity-0}
  we see that there exists a constant $C$ such that for all $\lambda \in [0,\lambda_0] $ 
\begin{align*} 
 & \left| R_{\psi_1,\psi_2}(\lambda;E\pm i \eta)    
-    \sum_{n=0}^{N-1} \lambda^n T_{n,\infty}[E \pm    i \eta ; \psi_1, \psi_2   ]     \right|    \\
& \leq  \left|  R_{\psi_1,\psi_2}(\lambda;E \pm i \eta)  
-    \sum_{n=0}^{  \lceil 4 (N+3)/\epsilon \rceil } \lambda^n T_{n,\infty}[E \pm    i \eta ; \psi_1, \psi_2]   \right|  +  \sum_{n=N}^{\lceil 4 (N+3)/\epsilon \rceil}   C_{(n)}  \lambda^n \nn  \\ 
&   \leq   C_0     \lambda^N  +   \sum_{n=N}^{\lceil 4 (N+3)/\epsilon \rceil}  C_{(n)}  \lambda^n  \leq C \lambda^{N}    .  \nn 
\end{align*}  
This shows the claim. 
\end{proof}

\begin{remark}{\rm 
We note that  Theorem \ref{asymptotictheorem} gives  an   asymptotic expansion 
of  $R_{\psi_1,\psi_2}(\lambda;E + i \eta)$ with   $\lambda$-dependent expansion coefficients, which are 
 continuous and bounded. Explicitly,  it follows by the triangle inequality  from  Theorems  \ref{asymptotictheorem} (dividing  \eqref{expansionasymptoticmain} for $N+1$  by $\lambda^N$)
 and  \ref{thm:boundaryvelueexpcoeff} that  
\begin{align} \label{asympprop} 
\lim_{\lambda \downarrow 0}\frac{  R_{\psi_1,\psi_2}(\lambda;E + i \eta) -\sum_{n=0}^{N-1} T_{n,\infty}[E+i\eta;\psi_1,\psi_2] \lambda^n }{  \lambda^{N}}  =   T_{N,\infty}[E+i 0+;\psi_1,\psi_2] . 
\end{align} 
Furthermore, it follows from \eqref{asympprop}, that whenever the right hand side  of  \eqref{asympprop} 
is nonzero, the functions 
$$
\left[ R_{\psi_1,\psi_2}(\lambda;E + i \eta) -\sum_{n=0}^{N-1}T_{n,\infty}[E+i\eta;\psi_1,\psi_2] \lambda^n \right]   \quad \text{ and } \quad 
T_{N,\infty}[E+i 0+;\psi_1,\psi_2]  \lambda^N 
$$
 are for $\lambda \downarrow 0$ asymptotically equivalent   according to the standard definition, cf.  \cite{deBruijn.1981}. 
In fact, the expansion \eqref{expansionasymptoticmain} given in the Theorem  \ref{asymptotictheorem} is for   $\lambda \downarrow 0$  a so called  extended asymptotic expansion with respect to the asymptotic sequence $( \lambda^n )_{n \in \N_0}$, cf. \cite{EstradaKanwal.2002}.

}
\end{remark}

\begin{remark}{\rm 
To obtain an  asymptotic expansion for  the integrated density of states analogous  to   Theorem \ref{asymptotictheorem}  one  has to  calculate   the normalized trace over $L^2(\Lambda_L)$ of the right hand side of  \eqref{defoflimit-00}. Proving  that this normalized trace is finite  requires involved estimates, which the authors intend  to  perform  in   a subsequent paper. 
}
\end{remark}

\section{Pairing Potential Labels}
\label{PPL}
In this section we derive formulas about expectations of products of the random potential, which will be used in the following 
sections.  The main  formulas   have been shown   in \cite{ErdosSalmhoferYau.2008a}.   
We  are going to work with partitions of the set $ \{1,2,...,n\}$ of natural numbers. We want to first give a short introduction to explain how these occur, as well as how they are technically handled.  In the definition of $ T_{n,L}$ given in \eqref{defoffinT} we are faced with the expectation  value of powers of the random potentials, i.e.,  terms of the form
\begin{align}
\label{expvCP}
{\bf E}_v^{\otimes M} \ {\bf E}_{y_L}^{\otimes M} \sum_{\gamma_1,...,\gamma_n = 1}^M \prod_{j=1}^n  v_{\gamma_j}.
\end{align}
We are therefore going to consider products of the random potentials $v_{\gamma}$ at the sites $y_{L,\gamma} \in \Lambda_L$. Note,  that for  a given configuration $\{ \gamma_1,...,\gamma_n \}$, different indices $j_1, \ldots , j_l$ may label the same sites, i.e. $\gamma_{j_1} = \ldots = \gamma_{j_l}$.  This case would produce a factor
\begin{align*}
\prod_{i=1}^l  v_{\gamma_{j_i}}=v_{\gamma_{j_1}}^l , 
\end{align*}
since all  $v_{\gamma_j}$  are the same. 
Thus  the expectation value of this factor yields  an $l$-th moment. 

To control this  technically, we introduce the following notation.
Let $\mathcal{A}_n$ be the set of partitions of $ \{1,2,...,n\}$.
For a given configuration $\{ \gamma_1,...,\gamma_n \}$ we are seeking to  filter for the unique partition $A$ that sorts the indicies $ 1,2,...,n$ in such a way, that the following two conditions  are satisfied. 
\begin{itemize}
\item[(a)] All indicies in each $a \in A$ correspond to the same site.
\item[(b)] All sets $a$ correspond to different sites.
\end{itemize}
To express this, we define for $A \in \mathcal{A}_n$ the two functions $\chi_A$ and $\tilde{\chi}_A$ on $\N^n$
as follows.   First,  we set  
\muun{
\label{X1}
\chi_A(z_1,....,z_n) := \prod_{a \in A} \bigg[ \prod_{(j,l) \in a \times a } 1_{\{  z_j = z_l \}} \bigg] ,
}
which ensures condition (a).   Second,  we define  
\muun{
\label{X2}
\tilde{\chi}_A(z_1,....,z_n) := \chi_A(z_1,...,z_n)  \prod_{\substack{ ( a, b )   \in A \times A :  \\ a \neq b}}  \bigg[ \prod_{(j,l) \in a \times b} 1_{\{  z_j \neq  z_l \}} \bigg]   ,
}
which ensures that condition (b) is additionally met.
Observe that for any $(z_1,...,z_n) \in \N^n$ we have 
\begin{align}
\label{upartition}
\sum_{A \in \mathcal{A}_n}  \tilde{\chi}(z_1,...,z_n)  = 1,
\end{align}
since there is exactly one unique partition that satisfies  both conditions (i) and (ii).
With this small preamble, we can now express the expected value in term (\ref{expvCP}) and show the following lemma.

\begin{lemma} \label{lem:combprob1} 
For any fixed $L > 0$,  integers  $n, M  \in \N$,  and any fixed momenta $q_j \in \Lambda_L^*$, $j=1,...,n$,  the identity 
\begin{align*}
 {\bf E}_v^{\otimes M} \ {\bf E}_y^{\otimes M} \sum_{\gamma_1,...,\gamma_n = 1}^M \prod_{j=1}^n  v_{\gamma_j} \exp(2 \pi i q_j \cdot y_{L,\gamma_j}) =  \sum_{A \in \mathcal{A}_n} 1_{\{ |A| \leq M  \}} \frac{M!}{(M-|A|)!}
\prod_{a \in A} \left\{ m_{|a|} \delta_* \left( \sum_{l \in a} q_l \right)  \right\} 
\end{align*} 
holds.
\end{lemma} 
\begin{proof} 
Let $A \in \mathcal{A}_n$ and let $\chi_A$ and $\tilde{\chi}_A$ be defined as  in \eqref{X1} and \eqref{X2} above.
Inserting the identity \eqref{upartition} into the left hand side we find 
\begin{align*}
 & {\bf E}_v^{\otimes M} \ {\bf E}_y^{\otimes M} \sum_{\gamma_1,...,\gamma_n = 1}^M \prod_{j=1}^n  v_{\gamma_j} \exp(2 \pi i q_j \cdot y_{L,\gamma_j}) \\
& = {\bf E}_v^{\otimes M} \ {\bf E}_y^{\otimes M} \sum_{\gamma_1,...,\gamma_k = 1}^M  \sum_{A \in \mathcal{A}_n}  \tilde{\chi}_A(\gamma_1,...,\gamma_n)  \prod_{j=1}^n  v_{\gamma_j} \exp(2 \pi i q_j \cdot y_{L,\gamma_j})\\
& = \sum_{A \in \mathcal{A}_n}   {\bf E}_v^{\otimes M} \ {\bf E}_y^{\otimes M} \sum_{\gamma_1,...,\gamma_n = 1}^M  \tilde{\chi}_A(\gamma_1,...,\gamma_k)  \prod_{j=1}^n  v_{\gamma_j} \exp(2 \pi i q_j  \cdot y_{L,\gamma_j})\\
& =
\sum_{A \in \mathcal{A}_n}   \sum_{\gamma_1,...,\gamma_n = 1}^M  \tilde{\chi}_A(\gamma_1,...,\gamma_n)  \prod_{a \in A} \bigg[  {\bf E}_v^{\otimes M} \prod_{j \in a}   v_{\gamma_j} \bigg] \left[ {\bf E}_y^{\otimes M}\exp \left(2 \pi i \sum_{j\in a} q_j \cdot y_{L,\gamma_j} \right) \right] \\
 & \overset{(*)}{=}\sum_{A \in \mathcal{A}_n}   \sum_{\gamma_1,...,\gamma_n = 1}^M  \tilde{\chi}_A(\gamma_1,...,\gamma_n) 
 \prod_{a \in A} m_{|a|}  \delta_* \left( \sum_{j \in a} q_j \right).
\end{align*} 
Note that $(*)$ follows from the discussion about the design of  $ \tilde{\chi}_A(\gamma_1,...,\gamma_n) $ prior to the lemma, as well as the fact that $A$ is a partition. The details are as follows.  Evaluating the  $\tilde{\chi}_A(\gamma_1,...,\gamma_n)$ ensures by \eqref{X1} that for a fixed $a \in A$ all indices $\gamma_j$ with $j \in a$ are the same index, hence all $y_{L,\gamma_j}$ with $j \in a$ are the same site.  
Therefore the $v_{\gamma_j}$ are the same random variables and we find
\muu{
{\bf E}_v^{\otimes M} \prod_{j \in a}   v_{\gamma_j} = m_{|a|}.
}
The definition of ${\bf E}_y^{\otimes M}$ in \eqref{esy:3.240} together with
\muu{
 \frac{1}{|\Lambda_L|} \int_{\Lambda_L} \exp \left(2 \pi i \sum_{j\in a} q_j \cdot y \right) dy = \delta_* \left( \sum_{j \in a} q_j \right)
}
finally yield $(*)$.
Now the claim follows from the combinatorical identity 
\begin{align*}
\sum_{\gamma_1,...,\gamma_n = 1}^M  \tilde{\chi}_A(\gamma_1,...,\gamma_n) 
&= 1_{|A| \leq M}  \frac{M!}{(M-|A|)! },
\end{align*}
which can be seen as follows. 
The LHS sums over all configurations of the $n$ indices that stand for $M$ possible points, while evaluating the  $\tilde{\chi}_A$-function for a given $A$.   In fact, since all $a \in A$ - which are $|A|$ many sets - correspond to different ones of the $M$ possible points each.  So for the first set $a_1$ there are  $M$ choices for the second one $(M-1)$ choices and so on and for the last set $a_{|A|}$ there are $(M-|A|+1)$ choices left.
 This is exactly expressed through the RHS which coincides with a $|A|$-permutation of $M$.  That is,  the number of ways of choosing $|A|$ out of $M$ elements without repetition where the order is taken in to account.  
\end{proof} 

\begin{lemma}\label{lemP}  For a Poisson distributed random variable  $N$ with mean $\lambda$ we  
 have for any $k \in \N$ 
\begin{align} 
\E \left[ \prod_{j=0}^{k-1} (N-j) \right]   
& = \lambda^k . \label{eq:poissonexp} 
\end{align} 
\end{lemma} 
The proof of Lemma \ref{lemP} is usually proven  using  the probability-generating function, see for example   \cite[Section 1.1]{LastPenrose.2018} or \cite{Kingman.1993}. 

 The following lemma is a continuation of Lemma \ref{lem:combprob1} in the sense that we evaluate the $\E_M$ expectation as well.

\begin{lemma} \label{lem:combprob2} Let $L > 0$,  $n \in \N$, and  $q_j \in \Lambda_L^*$, $j=1,...,n$.  Then
\begin{align*}
{\bf E}_M  {\bf E}_v^{\otimes M} \ {\bf E}_{y_L}^{\otimes M} \sum_{\gamma_1,...,\gamma_n = 1}^M \prod_{j=1}^n  v_{\gamma_j} \exp(2 \pi i q_j \cdot  y_{L,\gamma_j}) =  \sum_{A \in \mathcal{A}_n} 
\prod_{a \in A} \left\{ m_{|a|} \delta_{*,L} \left( \sum_{l \in a} q_l \right)  \right\} 
\end{align*}   
holds.
\end{lemma} 
\begin{proof}
The statement follows from \Cref{lem:combprob1} and the identity 
$$
{\bf E}_M   1_{|A| \leq M } \frac{M!}{(M-|A|)!} =  |\Lambda_L|^{|A|},
$$
which follows from  \Cref{eq:poissonexp}. 
\end{proof} 
To transform into momentum space we shall expand with respect to the ONB $\varphi_p$ defined in \eqref{eq:deofonb}. That is we  shall  insert the following  identity, which 
hold in strong operator topology 
\begin{equation} \label{fourierident} 
\psi= \sum_{p \in \Lambda_L^*}  \SP{\varphi_p ,  \psi}   \varphi_p   =  \int_{\Lambda_L^*}  \SP{e^{  2 \pi i \inn{ p  , \cdot } } , \psi (\cdot)} e^{  2 \pi i \inn{ p , \cdot}  } dp,
\end{equation} 
for all $\psi \in L^2 (\Lambda_L)$, which provides  the  representation via the ONB $\{\varphi_p : p \in \Lambda_L^* \}$. 

Moreover, for $\psi\in L^2 (\Lambda_L)$ and $ W \in L^\infty (\Lambda_L)$ (since $\Lambda_L$ is compact they are both also in $L^1 (\Lambda_L)$)  we will use the identities  
\begin{align} 
 \langle e^{  2 \pi i \inn{ p  , \cdot  } },  \psi \rangle & = \int_{\Lambda_L}  e^{ - 2 \pi i p \cdot x } \psi(x) dx = \widehat{\psi}(p) \label{ftident1}  \\
\langle e^{  2 \pi i \inn{ p , \cdot }  },   W(\cdot)   e^{  2 \pi i \inn{  q , \cdot } } \rangle & =   \int_{\Lambda_L}    W(x)   e^{ - i 2 \pi (p-q)\cdot x } dx  = \widehat{W}(p-q). \label{ftident2} 
\end{align} 
 First we use the probabilistic structure of the potential \eqref{bumpnotationres} and calculate  the Fourier transform   
\muun{ \label{bumpnotationresdd}   
\widehat{ V}_{L,\omega}(p) 
&=  \int_{\Lambda_L} \sum_{\gamma=1}^M  v_\gamma B_\#(x-y_{L,\gamma})  e^{  -2 \pi i p \cdot x } d x \nn \\
&= \sum_{\gamma=1}^M v_\gamma \int_{\Lambda_L}  B_\#(z)  e^{ - 2 \pi i p \cdot (z+y_{L,\gamma})} d z \nn \\
&=  \sum_{\gamma=1}^M  v_\gamma \widehat{B}_\#(p) e^{ -2 \pi  i p \cdot  y_{L,\gamma}} .
}
In the following sections we will make use of the following lemma. 
 
\begin{lemma} \label{expProd00123} For the model introduced in Section \ref{Model} we have 
\begin{align}
& { \bf E}_L \left(   \prod_{j=1}^n  \hat{V}_{L,\omega}(p_j - p_{j+1} )    \right)
=  \sum_{A \in \mathcal{A}_n}  \prod_{a \in A} \left\{ m_{|a|}  \delta_{*,L} \left( \sum_{l \in a} (p_l - p_{l+1}) \right) 
  \prod_{l \in a}  \hat{B}_\#(p_l - p_{l+1} ) \right\}  .  \label{eq:ofexpofrandpot2} 
\end{align} 

\end{lemma} 
\begin{proof} 
Inserting  \eqref{bumpnotationresdd} into the left hand side of \eqref{eq:ofexpofrandpot2} in the first, step multiplying out the product over the sum in the second step,  using  \Cref{lem:combprob2} in the final step we find
\begin{align*}
& { \bf E}_L \left(   \prod_{j=1}^n  \hat{V}_{L,\omega}(p_j - p_{j+1} )    \right)  \nonumber  \\
&= { \bf E}_L \left(   \prod_{j=1}^n \sum_{\gamma=1}^M  
  v_{\gamma}   \hat{B}_\#(p_j - p_{j+1} ) e^{ -2 \pi  i (p_j - p_{j+1} ) \cdot  y_{L,\gamma}}   \right)  \nonumber  \\
& =   { \bf E}_L \sum_{\gamma_1,...,\gamma_n =1}^M \prod_{j=1}^n  \left\{ 
  v_{\gamma_j}   \hat{B}_\#(p_j - p_{j+1} ) e^{ -2 \pi  i (p_j - p_{j+1} ) \cdot y_{L,\gamma_j}} \right\}  \nonumber 
  \\
 & =  \sum_{A \in \mathcal{A}_n}  \prod_{a \in A} \left\{ m_{|a|}  \delta_{*,L} \left( \sum_{l \in a} (p_l - p_{l+1}) \right) 
  \prod_{l \in a}  \hat{B}_\#(p_l - p_{l+1} ) \right\}  .  
\end{align*} 
\end{proof}

\section{Expectation of the  Expansion Coefficients} 

\label{sec:prooasymexp}

In this section, we prove Theorems   \ref{lem:formulaforexpcoeff},    \ref{exinflim},   and   \ref{thm:boundaryvelueexpcoeff}.
We introduce the notation for the canonical norm in the $\ell^q$-spaces.  For  $f \in \ell^q(\Lambda_L^*)$ 
with $q  \in [1,  \infty]$,
we write 
  \begin{align*} \| f \|_{*, q} = \left( \int_{\Lambda_L^*} |f(k)|^q dk \right)^{1/q}  \ ( 1 \leq q < \infty)   , \quad   \| f \|_{*, \infty } = \sup_{k \in \Lambda_L^*} |f(k)| .
  \end{align*}  

\begin{proof}[Proof of Theorem \ref{lem:formulaforexpcoeff}] We use  the definition given in  \eqref{defoffinT} and  
 transform into momentum  space by  means of the  identities in Eqns. \eqref{fourierident}--\eqref{ftident2} as well as  \eqref{eq:deoflaplac}.  
This gives for $z \in \C \setminus [0,\infty)$, where for reasons of lack of space we write $d p := d(p_1, \ldots , p_{n+1})$ 
\begin{align}\label{eq:expforTn0}  
& T_{n,L}[z;\psi_1,\psi_2]   \\
&= {\bf E}_L   \inn{ \psi_{1,\#} ,  R_L(z) [  V_L  R_L(z)  ]^n  \psi_{2,\#} }_L 
\nonumber \\
&  =  {\bf E}_M   {\bf E}_v^{\otimes M} {\bf E}_{y_L}^{\otimes M}    \int_{(\Lambda_L^*)^{n+1} }   \prod_{j=1}^n  \widehat{V}_{L,\omega}(p_j - p_{j+1} )  
\overline{\widehat{\psi}}_{1,\#}(p_1) \widehat{\psi}_{2,\#}({p}_{n+1})  \left(    \prod_{j=1}^{n+1}      \frac{1}{ \nu(p_j) - z}        \right)  d p \nonumber \\ 
&  = {\bf E}_M  \  \int_{(\Lambda_L^*)^{n+1} }  {\bf E}_v^{\otimes M} {\bf E}_{y_L}^{\otimes M}  \prod_{j=1}^n  \widehat{V}_{L,\omega}(p_j - p_{j+1} )  
\overline{\widehat{\psi}}_{1,\#}(p_1) \widehat{\psi}_{2,\#}({p}_{n+1})  \left(    \prod_{j=1}^{n+1}      \frac{1}{ \nu(p_j) - z}        \right)  d p \nonumber
\end{align}
where in the last identity we used Fubinis theorem, which we justify in the following paragraph.

For fixed $M$ we find  with  a canonical change of summation variables 
\begin{align}
&  \int_{(\Lambda_L^*)^{n+1} } \left|  \prod_{j=1}^n  \widehat{V}_{L,\omega}(p_j - p_{j+1} )  
\overline{\widehat{\psi}}_{1,\#}(p_1) \widehat{\psi}_{2,\#}({p}_{n+1})  \left(    \prod_{j=1}^{n+1}      \frac{1}{ \nu(p_j) - z}        \right)   \right|    d p \nonumber \\ 
&  \leq \frac{  \| \widehat{V}_{L,\omega} \|_{*,1}^n  \| \widehat{\psi}_{1,\#} \|_{*,2}   \| \widehat{\psi}_{2,\#} \|_{*,2}  }{{\rm dist}(z,[0,\infty))^{n+1}}    .
\end{align} 
Now using \eqref{bumpnotationresdd}   we find  
\begin{align} \label{VL1bound}  
\| \widehat{V}_{L,\omega} \|_{*,L,1} \leq   \sum_{j=1}^M |v_j|  \| \widehat{B}_\# \|_{*,1} .   
\end{align} 
By the symmetry assumptions on $B$ we will show in Lemma \ref{lemestfourdisc}, below, that $ \| \widehat{B}_\# \|_{*,1}$ is finite.
Any power of the right hand side of  \eqref{VL1bound}   is thus  integrable with respect to  ${\bf E}_v^{\otimes M} {\bf E}_{y_L}^{\otimes M} $  by means of  the moment assumption \eqref{2.4}. 
This justifies the use of Fubini's theorem and therefore in the last identity of  \eqref{eq:expforTn0}.

Now using  Lemma \ref{lem:combprob1} and Lemma  \ref{lemP} as well as Fubini's theorem we can interchange ${\bf E}_M$  and the  sum over  the momentum variables and find  

\begin{align*}
&  T_{n,L}[z;\psi_1,\psi_2]\\ 
&  =    \int_{(\Lambda_L^*)^{n+1} } {\bf E}_M  {\bf E}_v^{\otimes M} {\bf E}_{y_L}^{\otimes M}  \prod_{j=1}^n  \widehat{V}_{L,\omega}(p_j - p_{j+1} )   
\overline{\widehat{\psi}}_{1,\#}(p_1) \widehat{\psi}_{2,\#}({p}_{n+1}) \\
&\times  \left(    \prod_{j=1}^{n+1}      \frac{1}{ \nu(p_j) - z}        \right)  d (p_1 , \ldots p_{n+1})  \\
& =   \int_{(\Lambda_L^*)^{n+1}}   \sum_{A  \in \mathcal{A}_n} \mathcal{P}_{A,L}(k)   \overline{\widehat{\psi}}_{1,\#}(k_1)     \widehat{\psi}_{2,\#}(k_{n+1}) \prod_{j=1}^{n+1}  (\nu( k_j) - z)^{-1} d( k_1, \ldots, k_{n+1} ) ,  
\end{align*}
where in the last identity we used Lemma \ref{expProd00123} to insert \eqref{eq:ofexpofrandpot2}  and used  the definition given in  \eqref{eq:defofdeltapi0}. 
\end{proof}

Before we can show the other two theorems we will need some preparatory work. 
For this we introduce  the notation  
 \begin{align}
 \label{Klammern}
 \langle x \rangle = (1 + x^2)^{1/2},
 \end{align} which we shall use throughout the paper.
The following lemma will be applied in the proof of Proposition \ref{propexofcinf}.  In it's proof, we will be making use of Lemma \ref{ChSymbEst}.
\begin{lemma} \label{lemestfourdisc}    There exists a constant $C_d$ (explicitly given in the proof)
 and  a constant $\tilde{C}_d$ such that for all $L \geq 1$ and  $f \in \mathcal{S}(\R^d)$ which are 
symmetric  with respect to the reflections $\rho_j : (x_1,...,x_j,...,x_d) \mapsto (x_1,...,-x_j,...,x_d)$ for all $j=1,...,d$ (i.e.  $f \circ \rho_j = f$ for all $j=1,...,d$), 
we have 
\begin{align}  \label{eq:dgenallk0} 
\langle  k_1 \rangle^2 \cdots  \langle k_d \rangle^2 | \widehat{f}_\#(k) |  \leq C_d  \sum_{\substack{\alpha \in \N_0^d : \\  \alpha_j \leq 2}} \sup_{x \in \R^d}  |  \langle x \rangle^{2d} \partial^\alpha f(x) | 
 \end{align} 
for all $k=(k_1, \ldots , k_d) \in \R^d$.
Furthermore $\widehat{f}_\# \in \ell_*^1(\Lambda_L^*)$,  and 
 \begin{align} \label{secondRest0} 
\| \widehat{f}_\# \|_{*,1} \leq \tilde{C}_d \sum_{\substack{ \alpha \in \N_0^d :  \\ \alpha_j \leq 2}} \sup_{x \in \R^d}| \langle x \rangle^{2d} \partial^\alpha f(x) |  . \end{align} 
\end{lemma} 
\begin{proof} 
We  will show   \eqref{eq:dgenallk0} by induction over $d$. 
The integrability and the $\ell^1$-bound  \eqref{secondRest0}  will then follow by dividing by  $\langle  k_1 \rangle^2 \cdots  \langle k_d \rangle^2$  and  summing over  each of the components of $k$ separately. 

 First consider $d=1$.  Using Lemma \ref{ChSymbEst},  clearly, for $k \in \Z/L $ with $|k| \leq 1$, we find  
 \begin{align}  \label{eq:d=1small} 
 | \widehat{f}_\#(k)|  =\left|  \int_{-L/2}^{L/2} e^{ - i 2 \pi k x }  f(x) dx \right| 
 \leq  \int_{-L/2}^{L/2}   \underbrace{\left| e^{ - i 2 \pi k x }  \right| }_{\leq 1} | f(x) |dx 
  \leq \pi \sup_{x \in \R} | \langle x \rangle^2 f(x) | .
\end{align}
On the other hand  for $k \in \Z/L \setminus \{0 \}$ using integration by parts and the reflection symmetry \eqref{SC} we find
\begin{align*} 
k \widehat{f}_\#(k) &  = k  \int_{-L/2}^{L/2} e^{ - i  2 \pi k  x } f(x) dx  \\
&  =   \int_{-L/2}^{L/2} k  e^{ - i  2 \pi k  x } f(x) dx  \\
&  = \frac{1}{-i 2 \pi}   \int_{-L/2}^{L/2} \left( \frac{d}{dx}   e^{ - i  2 \pi k  x } \right) f(x) dx  \\
&  = \frac{1}{-i 2 \pi}   \left(   \underbrace{ [ e^{ - i  2 \pi k  x } f(x) ]_{-L/2}^{L/2}}_{=0}    - \int_{-L/2}^{L/2}  e^{ - i  2 \pi k  x } \frac{d}{dx} f(x) dx  \right) \\
&  = \frac{1}{2 \pi i }  \int_{-L/2}^{L/2}  e^{ - i  2 \pi k  x }  f'(x) dx  . 
\end{align*} 
Multiplying this by $k$, an analogous calculation using integration by parts yields 
\begin{align} 
\label{k2f}
k^2  \widehat{f}_\#(k) 
&  =\frac{1}{2 \pi i }     \int_{-L/2}^{L/2} k  e^{ - i  2 \pi k  x } f' (x) dx \nn  \\
&  =    \frac{1}{( 2 \pi)^2}   \int_{-L/2}^{L/2} \left( \frac{d}{dx}   e^{ - i  2 \pi k  x } \right) f'(x) dx  \nn \\
&  =   \frac{1}{( 2 \pi)^2}   \left(    [ e^{ - i  2 \pi k  x } f'(x) ]_{-L/2}^{L/2}   - \int_{-L/2}^{L/2}   e^{ - i  2 \pi k  x }  f''(x) dx  \right). 
\end{align} 
Using first the triangle inequality in \eqref{k2f} and then Lemma \ref{ChSymbEst} 
we find
\begin{align} 
k^2 | \widehat{f}_\#(k) | & \leq    \frac{1}{( 2 \pi)^2}\left(    2     \sup_{x \in \R}  |  \partial f(x) | + \int_{-L/2}^{L/2} | \partial^2 f(x) |  dx   \right) \nn \\
&  \leq c_1 \sum_{\alpha \in \N_0 : \alpha \leq 2 }  \sup_{x \in \R}  | \langle x \rangle^2 \partial^\alpha f(x) |, \label{eq:d=1klarge} 
\end{align} 
where $c_1 : = \frac{1}{2 \pi^2} + \frac{1}{4 \pi }$

Using $\langle k \rangle^2 \leq   2( 1_{|k| \geq 1} k^2 +  1_{|k| \leq 1} )$ and then  \eqref{eq:d=1klarge} for the first summand and   \eqref{eq:d=1small} for the second summand shows
\muun{
\label{kAufteilung}
\langle  k \rangle^2 | \widehat{f}_\#(k) |  
&\leq 2( 1_{|k| \geq 1} k^2 | \widehat{f}_\#(k) | +  1_{|k| \leq 1} | \widehat{f}_\#(k) |)  \nn
\\
& \leq \left( \frac{1}{\pi^2} + \frac{1}{2 \pi} + 2 \pi\right)   \sum_{\substack{\alpha \in \N_0 : \\  \alpha \leq 2}} \sup_{x \in \R}  |  \langle x \rangle^{2} \partial^\alpha f(x) |, 
}
 which is exactly \eqref{eq:dgenallk0} for $d=1$ with $C_1 := \frac{1}{\pi^2} + \frac{1}{2 \pi} + 2 \pi$.
Now let us asume   \eqref{eq:dgenallk0} holds for $d-1$ with some constant $C_{d-1}$ independent of $L$. We want to show that it also holds for $d$.  At this point,  we define the Fourier transform in the last variable by 
\begin{align}
\label{SchwFct}
\widehat{f}_\#(x_1,...,x_{d-1};k_d)  :=  \int_{-L/2}^{L/2}  e^{- 2 \pi i k_d   x_d } f(x_1, \ldots, x_d) dx_d .
\end{align}  
Let $k_d \in \Z/L$. First observe that the right hand side of \eqref{SchwFct}
is again a Schwartz function on $\R^{d-1}$  which is symmetric with respect to  $\rho_{j}$, $j=1,...,d-1$. 
Using    Fubini's theorem, we find
\begin{equation} \label{iteratedfourier} \widehat{f}_\#(k',k_d) = \int_{[-L/2,L/2]^{d-1}} \widehat{f}_\#(x';k_d) e^{- 2 \pi i k' \cdot x'} dx'. 
\end{equation}
So  for  $|k_d|\leq 1$  we find from first the induction hypothesis, then \eqref{iteratedfourier}, triangle inequality and finally Lemma \ref{ChSymbEst} applied to the function $x_d \mapsto \partial^{\alpha'}   f(x',x_d) $
that
\begin{align} 
& \langle  k_1 \rangle^2 \cdots  \langle k_d \rangle^2 | \widehat{f}_\#(k_1,...,k_{d-1}, k_d) | \nonumber \\
 &  \leq C_{d-1}  \sum_{\alpha' \in \N_0^{d-1} : \, \alpha'_j \leq 2} \sup_{x' \in \R^{d-1}} 
 \left|  \langle x' \rangle^{2(d-1)}  \partial^{\alpha'}  \widehat{f}_\#(x';k_d)  \right| \nonumber  \\
 &  = C_{d-1}  \sum_{\alpha' \in \N_0^{d-1} : \,  \alpha'_j \leq 2} \sup_{x' \in \R^{d-1}} 
 \left|  \langle x' \rangle^{2(d-1)}   \partial^{\alpha'} \int_{-L/2}^{L/2}\ e^{- 2 \pi i k_d   x_d }   f(x',x_d)dx_d  \right|  \nonumber \\
 &  \leq C_{d-1}  \sum_{\alpha' \in \N_0^{d-1} : \, \alpha'_j \leq 2} \sup_{x' \in \R^{d-1}}  \langle x' \rangle^{2(d-1)} 
\int_{-L/2}^{L/2} \underbrace{  \left|   e^{- 2 \pi i k_d   x_d } \right|}_{\leq 1}  \left|  \partial^{\alpha'}   f(x',x_d) \right| dx_d    \nonumber \\
&  \leq C_{d-1}  \sum_{\alpha' \in \N_0^{d-1} : \, \alpha'_j \leq 2} \sup_{x' \in \R^{d-1}}  \langle x' \rangle^{2(d-1)} 
\left( \pi \sup_{x_d \in \R} \left|    \langle x_d \rangle^{2} \partial^{\alpha'}   f(x',x_d) \right| \right) \nonumber \\
  &  \leq C_{d-1} \pi  \sum_{\alpha \in \N_0^{d} :\,  \alpha_j \leq 2} \sup_{x \in \R^{d}} 
 \left|  \langle x \rangle^{2d } \partial^{\alpha} f(x)  \right| . \label{eq:dg1small}
\end{align} 
On the other hand lets consider the estimate for any  $k_d \in \Z/L$ with  $k_d \neq 0$. Then an analogous calculation as for $d=1$ using integration by parts shows  
\begin{align*} 
k_d \widehat{f}_\#(x_1,...,x_{d-1};k_d) &  = k_d  \int_{-L/2}^{L/2} e^{ - i  2 \pi k_d  x_d } f(x) dx_d  \\
&  =   \int_{-L/2}^{L/2} k_d  e^{ - i  2 \pi k_d  x_d } f(x)  dx_d  \\
&  = \frac{1}{-i 2 \pi}   \int_{-L/2}^{L/2} \partial_{x_d}   e^{ - i  2 \pi k_d  x_d }  f(x) dx_d  \\
&  = \frac{1}{-i 2 \pi}   \left(   \underbrace{ [ e^{ - i  2 \pi k_d  x_d } f(x)  ]_{x_d=-L/2}^{x_d=L/2}}_{=0}    - \int_{-L/2}^{L/2}  e^{ - i  2 \pi k_d  x_d } \partial_{x_d} f(x)  dx_d  \right) \\
&  = \frac{1}{2 \pi i }  \int_{-L/2}^{L/2}  e^{ - i  2 \pi k_d  x_d }   \partial_{x_d} f(x)  dx_d  . 
\end{align*} 
Multiplying this by $k_d$ an analogous calculation using integration by parts gives 
\begin{align} 
\label{k2df}
 k_d^2  \widehat{f}_\#(x_1,...,x_{d-1}; k_d) 
&  =\frac{1}{2 \pi i }     \int_{-L/2}^{L/2} k_d  e^{ - i  2 \pi k_d  x_d } \partial_{x_d} f(x)  dx_d \nn \\
&  =    \frac{1}{ (2 \pi)^2}   \int_{-L/2}^{L/2} \left( \partial_{x_d}    e^{ - i  2 \pi k_d  x_d } \right) \left( \partial_{x_d} f(x) \right) dx_d \nn \\
&  =   G_L (x_1,...x_{d-1};k_d), 
\end{align} 
where we defined 
$$
 G_L (x_1,..., x_{d-1};k_d) :=  \frac{1}{( 2 \pi)^2}   [ e^{ - i  2 \pi k_d x_d } \partial_{x_d} f(x)  ]_{x_d=-L/2}^{x_d=L/2}   -  \frac{1}{( 2 \pi)^2}  \int_{-L/2}^{L/2}  e^{ - i  2 \pi k_d  x_d }   \partial_{x_d}^2 f(x) dx_d.
$$
Now observe that $G_L(\cdot; k_d)$ is  a Schwartz function on $\R^{d-1}$ with the property that it is symmetric with respect to the reflections $\rho_j$ with $j=1,...,d-1$.
Now  from \eqref{iteratedfourier} and  the induction hypothesis  we find using  \eqref{k2df}  as in   \eqref{eq:d=1klarge}, while Lemma \ref{ChSymbEst} applied to $x_d \mapsto \partial_{x_d} ^2 \partial^{\alpha'} f(x',x_d) $ that 
\begin{align} 
& \langle k_1 \rangle^2 \cdots  \langle k_{d-1} \rangle^2  k_d ^2  | \widehat{f}_\#(k) | \nonumber   \\
&  \leq C_{d-1} \sum_{\alpha' \in \N_0^{d-1} : \alpha'_j \leq 2 } \sup_{x' \in \R^{d-1} } | \langle x' \rangle^{2(d-1)} k_d^2 \partial^{\alpha'}   \widehat{f}_\#(x';k_d) | \nonumber  \\
&  \leq C_{d-1}  \sum_{\alpha' \in \N_0^{d-1} : \alpha'_j \leq 2 } \sup_{x' \in \R^{d-1} }  \langle x' \rangle^{2(d-1)}
\nonumber  \\
& \times 
 \frac{1}{(2\pi)^2} \bigg( 2  \sup_{x \in \R}  | \partial_{x_d}  \partial^{\alpha'} f  (x',x_d) | + \int_{-L/2}^{L/2}  | \partial_{x_d} ^2 \partial^{\alpha'} f(x',x_d) | dx_d      \bigg) 
 \nonumber  \\
 &  \leq C_{d-1}  \sum_{\alpha' \in \N_0^{d-1} : \alpha'_j \leq 2 } \sup_{x' \in \R^{d-1} }  \langle x' \rangle^{2(d-1)}    c_{1,1} \sum_{\alpha \in \N_0 : \alpha \leq 2 }  \sup_{x_d \in \R}  | \langle x_d \rangle^2 \partial^\alpha \partial^{\alpha'} f(x',x_d) | \nonumber
  \\
 &  \leq C_{d-1}c_{1,1} \sum_{\alpha \in \N_0^{d} : \alpha_j \leq 2 } \sup_{x \in \R^{d} } | \langle x \rangle^{2d } \partial^\alpha f(x)  | . \label{eq:dg1klarge}
\end{align} 
Finally \eqref{eq:dg1klarge} and  \eqref{eq:dg1small}  show,  as we have seen in \eqref{kAufteilung} before,  that \eqref{eq:dgenallk0} holds for $d$ with $C_d = C_{d-1} C_1$ 
(note $C_d = C_1^d$).  Finally observe that  \eqref{eq:dgenallk0}  now implies \eqref{secondRest0}.
\end{proof}

\begin{remark} {\rm 
We note that if the Schwartz function  $B$  has compact support or if it is symmetric with respect to reflections as in  Lemma \ref{lemestfourdisc}, then there exists an $L_0 \geq 1$ such that for all $L \geq L_0$ we have 
\begin{align}  \label{eq:dgenallk} 
\langle  k_1 \rangle^2 \cdots  \langle k_d \rangle^2 | \widehat{B}_\#(k) |  \leq C_d  \sum_{\substack{\alpha \in \N_0^d : \\  \alpha_j \leq 2}} \sup_{x \in \R^d}  |  \langle x \rangle^{2d} \partial^\alpha B(x) |  .
 \end{align} 
In particular  for $L \geq L_0$ we have $\widehat{B}_\# \in \ell_*^1(\Lambda_L^*)$,  and 
 \begin{align} \label{secondRest} 
\| \widehat{B}_\# \|_{*,1} \leq \tilde{C}_d \sum_{\substack{ \alpha \in \N_0^d :  \\ \alpha_j \leq 2}} \sup_{x \in \R^d}| \langle x \rangle^{2d} \partial^\alpha B(x) |  . 
\end{align} 
In the compact case \eqref{eq:dgenallk} and \eqref{secondRest} follow from choosing $L_0 \geq 1$ sufficiently large such that $\supp B \subset (-L_0/2,L_0/2)^d$ holds while using  well known properties of  the Fourier transform of Schwartz functions.    In the symmetric case  \eqref{eq:dgenallk} and \eqref{secondRest} follow  from Lemma \ref{lemestfourdisc}.}
\end{remark} 

 The remaining part of this section is devoted to the proofs  of  Theorems  \ref{exinflim} and   \ref{thm:boundaryvelueexpcoeff},
 which state that the infinite volume limit for each expansion coefficient exists and that for this  infinite volume limit we can take the limit  as the spectral 
 parameter approaches the positive real axis from the complex upper or lower half plane.  Therefore,  we can write 
 \begin{align} \label{TintermsofC} 
T_{n,L}[ \cdot  ; \cdot , \cdot ]  & = \sum_{A \in \mathcal{A}_n}  C_{n, A, L }[ \cdot; \cdot , \cdot  ]   ,  
\end{align} 
where we defined 
  for $A  \in \mathcal{A}_n $ and $z \in \C \setminus [0,\infty)$ 
\begin{align} \label{eq:defofCpi}
C_{n,A,L}[z; \psi_1, \psi_2] & :=    \int_{(\Lambda_L^*)^{n+1}}  \mathcal{P}_{A,L}(k)   \overline{ \widehat{\psi}}_{1,\#}(k_1)   
  \widehat{\psi}_{2,\#}(k_{n+1}) \prod_{j=1}^{n+1}  (\nu( k_j) - z )^{-1} d (k_1, \ldots,  k_{n+1} ). 
\end{align}
We will now consider  the limit $L \to \infty$.  In the following we shall assume that 
$$
z \in \C \setminus [0,\infty) . 
$$
Technically,  it is more convenient to first sum over the discrete delta functions and then take the limit $L \to \infty$, rather than taking  the limit first and then 
integrating out the delta distributions cf.  Remark \ref{remdeltdirac}.\\ 
We introduce a change of variables given by
\begin{align} \label{changevar1}
u_0  = k_1 ,  \qquad  u_{s} = k_{s+1} - k_s,  \quad s=1,...,n
\end{align}
to be able resolve the discrete delta functions one at a time.
Expressing the variables $k$  in terms of the variables $u = (u_0,  \ldots u_n) \in (\R^d)^{n+1}$ we find
$$
k_j = \sum_{l=0}^{j-1} u_l  ,\quad j=1,...,n+1. 
$$
Applying this change of variables produces 
\begin{align} 
 C_{n,A,L}[z ; \psi_1,\psi_2] & =   \int_{(\Lambda_L^*)^{n+1}}   \prod_{ a \in A} \left\{  \delta_{*,L} \left( \sum_{s  \in a } u_s  \right) \prod_{s \in a} \widehat{B}_\#(-u_s)  \right\} \overline{ \widehat{ \psi}}_{1,\#}(u_0)    \nonumber  \\ & \quad  \widehat{\psi}_{2,\#} \left( \sum_{l=0}^{n}u_l  \right)
\prod_{j=1}^{n+1}  \left( \nu \left( \sum_{l=0}^{j-1} u_l \right)  -z  \right)^{-1} d (u_0,  \ldots, u_n).\label{eq:defofCpi}
\end{align}
For a given partition $A$ (hence given $a \in A$) we are now going to introduce a function $M_A$ to express the dependence of indicies belonging to the same $a \in A$ which is caused by the term 
$$
\delta_{*,L}\left( \sum_{s  \in a } u_s  \right). 
$$
This term  vanishes unless  $\sum_{s  \in a } u_s  =0$, which means that for every $a \in A$ we can express one of the variables by the negative sum of the others,  or  in the case,  where $a$ consists of only one element,  that element has to be equal to $0$.    
For each set $a \in A$ we choose to express the variable with index equal to  the highest number of  $a$
 by the negative sum of all the variables with index equal to any of  the other numbers in $a$, which gives us 
\begin{align*}
u_{\max a}=- \sum_{j \in a \setminus \{ \max a \} } u_j .
\end{align*}
  In the case where $a$ consists of only one element,  $ a \setminus \{ \max a \}= \emptyset$ which produces an empty sum, which is $0$ by convention.
To make the notation more usable we define the set of all  indices, which are the maximum of a set $a$ by
\begin{align}
J_A := \{ \max a : a \in A\}
\end{align}
as well as its complement
\begin{align}
I_A := \{ 1 ,  \ldots ,  n \} \setminus J_A.
\end{align}
Since $A$ is a partition, for any $j \in \{1 , \ldots , n\}$ there is a unique set $a \in A$ such that $j \in a$, we denote this set by $a (j)$. 
We define the map $M_A :( \R^d)^{| I_A |} \to (\R^d)^n$ as 
\begin{align}\label{changevar1next}
[M_A (v)]_j :=
\begin{cases}
v_j &: j \in I_A
\\
- \sum\limits_{ l \in a (j)  \setminus \{ j \} }  v_l &: j \in J_A,
\end{cases}
\end{align}
where $v=(v_1,  \ldots , v_n)$ with $v_j \in \R^d$ for all $j \in \{1, \ldots , n \}$. 
Note that \eqref{changevar1next} contains the case in which $j$ is the only element of $a(j)$ and  $[M_A (v)]_j =0$ holds.
This definition implies the following lemma.

\begin{lemma} \label{sum0}
Let $A$ be a partition of  $\{ 1 ,  \ldots ,  n \}$  and $M_A$ be defined as in  \eqref{changevar1next}.
Then we have
\begin{align}\label{totsumzero} 
\sum_{j=1}^n  [M_A (v)]_j = 0
\end{align}
for all $v  \in ( \R^d)^{| I_A |}$. 
\end{lemma}

\begin{proof}
This can easily be shown by inserting the definition
\begin{align*}
\sum_{j=1}^n  [M_A (v)]_j 
&= \sum_{j \in I_A}   [M_A (v)]_j +  \sum_{j \in J_A}   [M_A (v)]_j 
\\
&=\sum_{j \in I_A} v_j  + \sum_{j \in J_A} \left(  - \sum_{ l \in a (j)  \setminus \{ j \} }  v_l \right)
\\
&=\sum_{j \in I_A} v_j  - \sum_{j \in J_A}  \sum_{ l \in a (j)  \setminus \{ j \} }  v_l
\\
&\overset{(*)}{=}\sum_{j \in I_A} v_j  - \sum_{l \in I_A} v_l 
\\
&=0.
\end{align*}
It remains to show $(*)$. First observe that by definition it is straightforward to see that the  sets $a(j) \setminus \{ j \}$ for $j \in J_A$ are disjoint.   
Therefore it is sufficient to show the equality of the sets to obtain the equality of the sums.
Second, since by definition every $a \in A$ contains exactly one element of $J_A$,  while $A$ is a partition we now have
\begin{align*}
\bigcup_{j \in J_A} a(j) =  \bigcup_{a \in A} a = \{ 1 ,  \ldots ,  n \} .
\end{align*}
Since for every $j \in a(j)$ it now follows that 
\begin{align*}
\bigcup_{j \in J_A} \left( a(j)\setminus \{ j\} \right)  
=   \{ 1 ,  \ldots ,  n \} \setminus J_A =I_A.
\end{align*}
This  shows $(*)$ and ends the proof.
\end{proof}

With the notation  introduced  \eqref{changevar1next} summing in  \eqref{eq:defofCpi} over all  the variables  $u_j$ with $j \in J_A$  and evaluating the delta function yields, 
\begin{align} \label{eq:defofCpi2}
C_{n,A,L}[z ; \psi_1 ,\psi_2]  & =   \int_{\Lambda_L^*} du_0 \prod_{l \in I_A} \left( \int_{\Lambda_L^*}  dv_l \right)   \prod_{j  = 1}^n   \widehat{B}_\#(-[M_A(v)]_j)   
 \overline{ \widehat{\psi}}_{1,\#}(u_0)      \\ & \quad       
\widehat{\psi}_{2,\#}
\left( u_0  \right)   
 \prod_{j=1}^{n+1}  \left( \nu \left( u_0 +   \sum_{l=1}^{j-1} [M_A v]_l  \right) - z  \right)^{-1}  , \nonumber
\end{align}
where for the argument of $\widehat{\psi}_{2,\#}$ we used that by  Lemma \ref{sum0}  $u_0 + \sum_{l=1}^{n+1} [M_A v]_l = u_0$. 
Here and henceforth we adopt a notation where we write the integration variables right after the integral sign for notational compactness.
In Proposition \ref{propexofcinf},  we  will show  that  the limit $L \to \infty $ of   \eqref{eq:defofCpi2} exists.  To prove this, we will 
use results from the following lemma. 
For any $f \in L^1(\R^d)$ we define the Fourier series 
$$
\hat{f}(p) = \int_{\R^d} e^{ - 2 \pi i p \cdot x}  f(x) dx \text{ for } p \in \R^d  . 
$$
\begin{lemma} \label{fourierseries}  The following holds. 
\begin{enumerate}[(a)]
\item 
Let $f , g \in L^2(\R^d)$, then 
$$
\int_{\Lambda_L^*} \overline{\widehat{f}}_\#(k) \widehat{g}_\#(k) dk =  \int_{\Lambda_L} \overline{ f(x)} g(x) dx  
$$
\item 
If $f \in L^1(\R^d)$, then  
$$
\| \hat{f}_\# \|_{*,\infty} \leq \int_{\Lambda_L}  |f(x)|  dx \leq \| f \|_1 .  
$$
\item 
 \label{discconvgarten} If  $f \in L^1(\R^d)$,  then 
$$
\| \hat{f}_\# - \hat{f} \|_{*,\infty} \to 0 \quad ( L \to \infty ) . 
$$

\end{enumerate} 
\end{lemma}

\begin{proof} (a) is simply  the Parseval theorem for Fourier series. For details we refer the reader to \cite[Proposition 5.30 and Theorem 8.20]{folland}.  \\
(b) follows from the triangle inequality for integrals 
\begin{align*}
| \hat{f}_\#(k) |  &  = \left| \int_{\Lambda_L}  e^{-i2 \pi k \cdot x} f(x) dx  \right| \leq  \int_{\Lambda_L}   |  f(x) | dx  \leq \| f \|_1 .  
\end{align*} 
To prove (c),
we use 
\begin{align*}
| \hat{f}_\#(k) - \hat{f}(k) | &  = \left| \int_{\R^d} (1_{\Lambda_L} (x) - 1 ) e^{-i2 \pi k \cdot x} f(x) dx  \right|  \leq  \int_{\R^d} | (1 - 1_{\Lambda_L} (x)  )  f(x) | dx   . 
\end{align*} 
Now the right hand side tends to zero by the dominated convergence theorem.  
\end{proof} 

 In view of  \eqref{TintermsofC}, Theorem \ref{exinflim} will follow 
 from the next proposition, which shows the existence of the pointwise limit
 \begin{align} \label{TintermsofC2} 
\lim_{L \to \infty} T_{n,L}[ z   ; \cdot , \cdot ]  & = \sum_{A \in \mathcal{A}_n}  \lim_{L \to \infty}  C_{n, A, L }[ z ; \cdot , \cdot  ]  
\end{align} 
for $z \in \C \setminus [0,\infty)$.
\begin{proposition} \label{propexofcinf} 
For $z \in \C \setminus [0,\infty)$ and $\psi_1, \psi_2 \in L^2(\R^d)$ we have,  using the integral-notation introduced in \eqref{eq:defofCpi2} 
\begin{align} \label{eq:exoflimit2}
&   \lim_{L \to \infty} C_{n,A,L}[z; \psi_1, \psi_2]   \\
& =   \int_{\R^d} du_0 \prod_{l \in I_A} \left( \int_{\R^d}  dv_l \right)   \prod_{j  = 1}^n   \widehat{B}(-[M_A(v)]_j)  \overline{   \widehat{\psi}}_1(u_0)     \widehat{\psi}_2 \left( u_0\right)  \nn  \\
& \quad \prod_{j=1}^{n+1}  \left( \nu \left( u_0 +   \sum_{l=1}^{j-1} [M_A (v)]_l  \right) - z \right)^{-1}  ,  \nonumber
\end{align}
where $M_A$ is defined in  \eqref{changevar1next}.  For every  $n \in \N$, there exists a constant $K_n$ such that,  for all $L \geq 1$, $z \in \C \setminus [0,\infty)$, and  $\psi_1, \psi_2 \in L^2(\R^d)$,  we have 
\begin{align} \label{boundonCs} 
| C_{n,A,L}[z;\psi_1,\psi_2 ] | \leq \frac{ K_n   \| \psi_1 \| \| \psi_2 \|}{ {\rm dist}(z,[0,\infty))^{n+1}   }  . 
\end{align} 
\end{proposition} 

\begin{proof} \quad \\ 
\noindent
\underline{Step 1:} $C_{n,A,L}[z; \cdot , \cdot ]$ is a  sesquilinear form, which satisfies the bound  \eqref{boundonCs}.  

\vspace{0.25cm} 

\noindent
Estimating \eqref{eq:defofCpi2},    using \eqref{changevar1next},   Lemma \ref{lemestfourdisc}, and the elementary bound \eqref{ln-1},  we find   
that
\begin{align}
& | C_{n,A,L}[z ; \psi_1 ,\psi_2]| \nonumber \\
 & \leq    \int_{\Lambda_L^*} du_0 \prod_{l \in I_A} \left( \int_{\Lambda_L^*}  dv_l \right)   \prod_{j \in I_A} 
 \left| \widehat{B}_\#(-v_j)  \right|  \| \widehat{B}_\# \|_\infty^{n-|I_A|}     
|  \overline{ \widehat{\psi}}_{1,\#}(u_0)  |  \nonumber   \\ 
&\times  \left| \widehat{\psi}_{2,\#}
\left( u_0  \right)  \right| {\rm dist}(z,[0,\infty))^{-n-1}  \nonumber \\
& \leq   \| \widehat{B}_\# \|_{*,\infty}^{n-|I_A|}     \| \widehat{B}_\# \|_{*,1}^{|I_A|} \| \widehat{\psi}_{1,\#} \|_{*,2}  \| \widehat{\psi}_{2,\#} \|_{*,2}   {\rm dist}(z,[0,\infty))^{-n-1}  \nonumber \\
& \leq   \|B\|_1^{n-|I_A|}    C_B^{|I_A|} \| {\psi}_{1} \|_2  \| {\psi}_{2} \|_2   {\rm dist}(z,[0,\infty))^{-n-1}, \nonumber
\end{align}
where in the last inequality we used  Lemma  \ref{fourierseries} and  that   $\| \widehat{B}_\# \|_{*,1}$ can be 
bounded by a constant $C_B$,  which is independent of $L \geq 1$ by   Lemma \ref{lemestfourdisc}. 

\vspace{0.25cm}

\noindent
\underline{Step 2:} \eqref{eq:exoflimit2} holds for  $\psi_1, \psi_2 \in C_c^\infty(\R^d)$.  

\vspace{0.25cm}

This will be shown using the dominated convergence theorem writing the sum as an integral over  simple functions.
First choose $L$ sufficiently large such that $\supp \psi_j \subset \Lambda_L$. Then $\widehat{\psi}_{j,\#} = \widehat{\psi}_j |_{\Lambda_L^*}$.  Since the Fourier transform maps $C_c^\infty(\R^d)$ to Schwartz functions
$\widehat{\psi}_{j,\#} $ is the restriction of a Schwartz function.   For a function $f : (\Lambda_L^*)^{m} \to \C$, $m \in \N$ we define the function $E[f] :  (\R^d)^{m}  \to \C$ by $E[f](x) = f(k)$ if 
$x_j = k_j + \xi_j$ for $k_j \in \Lambda_L^*$ and $\xi_j \in [-\frac{1}{2L}, \frac{1}{2L})^d$, $j=1,...,m$.  With this definition we can express   \eqref{eq:defofCpi2} as an integral over simple functions. That is with    $m=1+ |I_A|$ and  $v = (v_1,....,v_{|I_A|})$ we find that 
\begin{align} \label{eq:defofCpi3}
C_{n,A,L}[z ; \psi_1 ,\psi_2]  & =   \int_{\R^d}  du_0 \prod_{l \in I_A} \left( \int_{\R^d}  dv_l \right)   \prod_{j  = 1}^n   E [\widehat{B}_\#(-[M_A( \cdot )]_j)](v)   
 E[\overline{ \widehat{\psi}}_{1}](u_0)      \\ &  \times       
E[\widehat{\psi}_{2}] ( u_0 )    E[Q](u_0,v) , \nonumber
\end{align}
with
$$
Q(u_0,v)  :=   \prod_{j=1}^{n+1}   \left( \nu \left( u +   \sum_{l=1}^{j-1} [M_A v]_l  \right)  - z  \right)^{-1}.
$$
First, we show that as $L \to \infty$ the integrand of \eqref{eq:defofCpi3} converges  to the integrand of  \eqref{eq:exoflimit2} pointwise.  To see this, observe that $Q$ is 
continuous as well as  $\hat{\psi_{1}}$ and $\hat{\psi_{2}}$  are continuous. Moreover,  from Lemma \ref{fourierseries} (c) and the continuity of $[M_A(\cdot)]_j$
 we see that $E [\widehat{B}_\#(-[M_A( \cdot )]_j)]$ converges  to $\widehat{B}_\#(-[M_A( \cdot )]_j) $ pointwise.  Finally, observe that the absolute value of the  integrand  of \eqref{eq:defofCpi3} is bounded  by a constant $C_{\psi_1,\psi_2,B,d,n}$ 
  uniformly for  $L \geq 1$ by 
\begin{align}
 \label{eq:estobmaj}  
& \left| \prod_{j  = 1}^n   E [\widehat{B}_\#(-[M_A( \cdot )]_j)](v)   
 E[\overline{ \widehat{\psi}}_{1}](u_0)     
E[\widehat{\psi}_{2}] ( u_0 )    E[Q](u_0,v) \right|  \nn    \\
& \leq
\|{B}\|_1^{n-|I_A|}   \prod_{j \in I_A}  \left\{ c_B E[ \langle v_j \rangle^{-2d} ] \right\} C_{\psi_1,\psi_2}E [  \langle u_0 \rangle^{-d-1} ]   {\rm dist}(z,[0,\infty))^{-n-1} \nn \\
& 
\leq C_{\psi_1,\psi_2, B,d,n}  \|{B}\|_1^{n-|I_A|}   \prod_{j \in I_A}  \left\{\langle v_j \rangle^{-2d} \right\}  \langle u_0 \rangle^{-d-1}   {\rm dist}(z,[0,\infty))^{-n-1} , 
\end{align}
where in the second inequality we used Lemma \ref{fourierseries} (b) to bound the terms involving $B$ for $j \in J_A$,  we used Lemma \ref{lemestfourdisc}
 to bound the terms involving $B$ for $j \in I_A$ with  some constant $c_B$,  we bounded the terms involving the wavefunctions $\psi_j$, $j=1,2$,  by a constant  $C_{\psi_1,\psi_2}$ (which can be done since $\hat{\psi}_1$ and $\hat{\psi}_2$  are  Schwartz functions), and for the 
 last factor  we used    \eqref{ln-1}. Note the  last inequality in  \eqref{eq:estobmaj}   follows from a simple squaring inequality. 
Now \eqref{eq:estobmaj}  is integrable with respect to  the measure $du_0 \prod_{j \in I_A} d v_j $.
Now applying the dominated convergence theorem completes Step 2.

\vspace{0.25cm} 

\noindent
\underline{Step 3:} \eqref{eq:exoflimit2} holds for  $\psi_1, \psi_2 \in L^2(\R^d)$.  

\vspace{0.25cm} 

\noindent
First, we observe that the right hand side of \eqref{eq:exoflimit2}     is a bounded sesquilinear form, which we call $h$. To see this, observe that on $C_c^\infty(\R^d)$ 
it is bounded as a limit (by Step 2)  of uniformly bounded forms (Step 1).   Thus,  for   $\psi_1', \psi_2' \in C_c^\infty(\R^d)$ from the triangle inequality we find 
\begin{align} 
&  | C_{n,A,L}[z; \psi_1, \psi_2]  - h(\psi_1,\psi_2) |  \label{trivbilinest} \\
&  \leq   | C_{n,A,L}[z; \psi_1, \psi_2 - \psi_2']  + C_{n,A,L}[z; \psi_1 - \psi_1',  \psi_2']| 
  \nn \\ & + |  C_{n,A,L}[z;\psi_1',\psi_2'] - h(\psi_1',\psi_2') | 
 +  | h(\psi_1'-\psi_1,\psi_2') - h(\psi_1,\psi_2 - \psi_2') | . \nn 
\end{align} 
Now the right hand side of \eqref{trivbilinest}  can be made arbitrarily small by means of  Step 1 and Step 2 by 
choosing first $\psi_j'$ sufficiently close to $\psi_j$ (by density) and then $L$ sufficiently large, respectively. 
\end{proof} 

In view of the above Proposition    \ref{propexofcinf}  we shall henceforth  write for $z \in \C \setminus [0,\infty)$ and $\psi_1, \psi_2 \in L^2(\R^d)$ 
\begin{align} \label{eq:exoflimit} 
& C_{n,A, \infty }[z; \psi_1, \psi_2]  :=  \lim_{L \to \infty} C_{n,A,L}[z; \psi_1, \psi_2]   .
\end{align}

As an immediate consequence of Proposition  \ref{propexofcinf} and the   coordinate transformation \eqref{changevar1} back to  the original
 variables, we obtain the  result of the following corollary, which is interesting of its own but  will not be needed in the sequel.

\begin{corollary}\label{remdeltdirac} For $z \in \C \setminus [0,\infty)$ and $\psi_1, \psi_2 \in \mathcal{S}(\R^d)$, we have 
\begin{align} \label{eq:exoflimit3}
& C_{n,A, \infty }[z ; \psi_1, \psi_2] \\
& =  \int_{(\R^d)^{n+1}}   \mathcal{P}_{A,\infty}(k) \overline{  \widehat{ \psi}}_1(k_1)     \widehat{\psi}_2(k_{n+1}) \prod_{j=1}^{n+1}  (\nu( k_j) - z )^{-1} d(k_1, \cdots,  k_{n+1})  , \nn 
\end{align} 
where 
\begin{align} \label{eq:defofdeltainf}
\mathcal{P}_{A,\infty}(k) & :=  \prod_{a \in A} \left\{ m_{|a|}  \delta \left( \sum_{l \in a} (k_l - k_{l+1}) \right) 
  \prod_{l \in a}  \widehat{B}(k_l - k_{l+1} ) \right\}  .
 \end{align}
  Here $\delta$ denotes the usual  Dirac measure  at the origin in $\R^d$.    
  \end{corollary} 
  
 In the remaining part of this section we prove Theorem  \ref{thm:boundaryvelueexpcoeff}. 
Henceforth,  we assume that the spectral parameter is of the form 
$z = E + i \eta$ with $$
E > 0  \text{ and } \eta > 0  ,
$$
the case $z = E - i \eta$ will follow by taking complex conjugates. 
Moreover we assume that   Hypothesis \ref{H1} holds and  that $\psi_1, \psi_2 \in L^2(\R^d)$ satisfy  Hypothesis \ref{H2}.


The main idea of the proof is to use a so called analytic dilation (see for example \cite{CombesThomas.1973,SR4}) to control the limit as $\eta \downarrow 0$.  For
 \eqref{eq:exoflimit}  using the 
right hand side of \eqref{eq:exoflimit2}  and  the change of variables $u_0 \mapsto e^{\theta} u_0$ and  $v_j \mapsto e^{\theta} v_j$ for $\theta \in \R$,  we find that 
\begin{align} \label{eq:defofCpi01}
& C_{n,A,\infty}[E + i \eta ; \psi_1 , \psi_2]  \\
& =   e^{ - d ( |I_A|+1 )\theta}  \int_{(\R^d)^{|I_A|+1}} du_0  \prod_{l \in I_A}  dv_l   \prod_{j = 1}^n  \widehat{B}(- e^{ -  \theta} [M_A(v)]_j )  \nonumber  \\
 & \quad \times  \overline{  \widehat{ \psi}}_1(e^{ -   \theta}u_0)     \widehat{\psi}_2 \left( e^{ -   \theta}  u_0 \right)  \prod_{j=1}^{n+1}  \left(e^{ - 2  \theta}  \nu \left( u_0 +   \sum_{l=1}^{j-1} [M_A v]_l  \right)  - E -  i \eta \right)^{-1}  . \nonumber
\end{align}
Note that the left hand side of \eqref{eq:defofCpi01} is independent of $\theta \in \R$.  By Hypothesis \ref{H1}, the assumption that $\psi_1$ and $\psi_2$ satisfy  Hypothesis \ref{H2} and the argument presented below, we will be   able to analytically extend the right hand side for  $\theta$ into  the  open set  
$$S := \left\{ \theta  \in \C :  - \frac{1}{2} \arctan(\eta/E)  <   {\rm Im} \theta <  3 \pi / 4  \right\} \cap \left\{ \theta \in \C : |  \theta | < {\rm min}(\vartheta_{\rm B},\vartheta_{\psi})  \right\},$$  
which is 
a  neighborhood of zero. As displayed in the Figure \ref{PictureDilatation} below, for $E,\eta > 0$, the point $E + i \eta$ is located in the first quadrant. At the same time $\nu \left( u_0 +   \sum_{l=1}^{j-1} [M_A (v)]_l  \right)$ is non-negative real-valued (for any $j=1, \ldots , n+1$), hence located on the non-negative real axis and gets turned by the factor $e^{ - 2  \theta}$ in negative direction by the angle $2 {\rm Im} \theta$. Since $\arctan(\eta/E)$  represents the angle of the polar form  of $E+i \eta$ and a rotation by $-\frac{3 \pi}{2}$ marks the transition between second and first quadrant,  
the first set in the intersection in the definition of $S$  ensures the invertibility of 
\begin{align} 
\label{TermPictureDilatation}
\left(e^{ - 2  \theta} \nu  \left( u_0 +   \sum_{l=1}^{j-1} [M_A v]_l  \right)  - E -  i \eta \right)
\end{align}
for any $j=1, \ldots , n+1$.

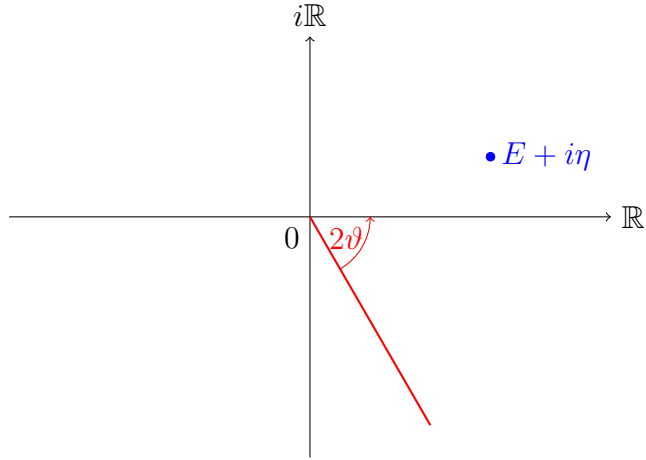
\begin{figure}[H]
\begin{center}

\begin{tikzpicture}[scale=0.8]
  \draw[->] (-5, 0) -- (5, 0) node[right] {$\R$};
  \draw[->] (0,  - 4) -- (0,3) node[above] {$i \R$};
  \draw[ red, thick] (0, 0) -- (2,-2*3^0.5) ;  
\filldraw[blue]  (3,1)  circle[radius=2pt] node[right] {$E+i\eta$};
\draw  (-0.3,0)   node[below] {$0$};
     \draw[red, <-] (1,0) arc (0:-60:1) ;
         \draw[red] (0.6,-0.35)  node {$ 2 \vartheta$};
\end{tikzpicture}

\caption{\label{PictureDilatation}  The set $\{  e^{  2 \theta} r  : r \geq 0 \}$  for $\vartheta = {\rm Im} \theta$ is drawn in red. }
\end{center} 
\end{figure}

To estimate the resolvents we will use the following lemma. 

\begin{lemma} Let   $\alpha, E , \eta, \lambda  \in \R$. Then    the following holds. 
\begin{itemize} 
\item[(a)] We have  
\begin{align} \label{estonres}  
|  e^{ - i \alpha } \lambda - E - i \eta  |  
&  =  | \lambda - e^{ i \alpha } ( E + i \eta) |   \geq  |\sin(\alpha) E + \cos(\alpha) \eta |.
\end{align} 
\item[(b)]  If $\alpha \in [0,\pi/2]$ and $\eta, E \in [ 0, \infty) $ then 
\begin{align} \label{estonresv2}  
|  e^{ - i \alpha } \lambda - E - i \eta  |  
 &  \geq  |\sin(\alpha) E|. 
\end{align} 
\end{itemize} 
\end{lemma} 
\begin{proof} 
(a) follows from 
\begin{align}  \label{estonres2} 
|  e^{ - i \alpha } \lambda - E - i \eta  |  
&  =  | \lambda - e^{ i \alpha } ( E + i \eta) | \nonumber  \\
&  = \left( (\lambda - \cos(\alpha) E + \sin(\alpha) \eta)^2 +   (\sin(\alpha) E + \cos(\alpha) \eta )^2\right)^{1/2} 
\end{align} 
and the monotonicity of the square root. 
Now (b) follows directly from (a). 
\end{proof} 
We are now ready to prove Theorem \ref{thm:boundaryvelueexpcoeff}.
\begin{proof}[Proof of Theorem \ref{thm:boundaryvelueexpcoeff}]  Let  \begin{align} \label{anaastheta} \vartheta \in (0,{\rm min}(\vartheta_{B},\vartheta_{\psi},3\pi/4)) . 
\end{align}
Then $i\vartheta \in S$. Let 
 $m = |I_A|$ and let a single $\int$  be  a short hand notation  for the collection of all the respective integrals.
From  \eqref{eq:defofCpi01} we find 
\begin{align} \label{eq:defofCpi1}
& C_{n,A,\infty }[E + i \eta; \psi_1 , \psi_2]  \\
& =   e^{ -  i d ( m+1 )\vartheta}  \int du_0  \prod_{s \in I_A}  dv_s   \prod_{j = 1}^n  \widehat{B}(- e^{ - i \vartheta} [M_A(v)]_j )    \overline{ \widehat{ \psi}}_1(e^{   i \vartheta}u_0)     \widehat{\psi}_2 \left( e^{ -  i \vartheta} u_0   \right) \nonumber  \\
& \quad  \prod_{j=1}^{n+1}  \left(e^{ - 2 i \vartheta} \nu \left( u_0 +   \sum_{l=1}^{j-1} [M_A (v)]_l  \right)  - E -  i \eta \right)^{-1}  . \nonumber
\end{align}
Factoring out $e^{ - 2 i \vartheta}$ in  the denominator, we find 
\begin{align} \label{eq:defofCpi222}
 & C_{n,A,\infty }[E + i \eta; \psi_1 , \psi_2] \\
& =  e^{ 2 (  n+1 ) i \vartheta} e^{- i d ( m+1 )\vartheta}  \int du_0  \prod_{s \in I_A}  dv_s   \prod_{j = 1}^n  \widehat{B}(- e^{ - i \vartheta} [M_A(v)]_j )    \overline{ \widehat{ \psi}}_1(e^{   i \vartheta}u_0)     \widehat{\psi}_2\left( e^{ -  i \vartheta} u_0   \right)  \nonumber  \\
& \quad   \prod_{j=1}^{n+1}  \left( \nu  \left( u_0 +   \sum_{l=1}^{j-1} [M_A v]_l  \right)  - e^{  2 i \vartheta} ( E +   i \eta )  \right)^{-1} 
 . \nonumber
\end{align}
Using  assumption    \eqref{anaastheta},
we can bound the integrand in  \eqref{eq:defofCpi222}  for   $\vartheta \in (0,\pi/4)$  by means of      \eqref{estonresv2}
\begin{align}
& \left| \prod_{j = 1}^n  \widehat{B}(- e^{ - i \vartheta} [M_A(v)]_j )    \overline{ \widehat{ \psi}}_1(e^{   i \vartheta}u_0)     \widehat{\psi}_2 \left( e^{ -  i \vartheta} u_0   \right)
  \prod_{j=1}^{n+1}  \left( \nu  \left( u_0 +   \sum_{l=1}^{j-1} [M_A v]_l  \right)  - e^{  2 i \vartheta} ( E +   i \eta )  \right)^{-1} 
\right|  \nonumber  \\
&  \leq \prod_{s \in I_A}     | \widehat{B}(- e^{ - i \vartheta} v_s) |  \| \widehat{B}(- e^{- i \vartheta } \cdot  )\|_\infty^{|A|}  \frac{  \left| \overline{ \widehat{ \psi}}_1(e^{   i \vartheta}u_0)     \widehat{\psi}_2 \left( e^{ -  i \vartheta} u_0   \right) \right|}{  |\sin(2 \vartheta ) E  |^{n+1} }   \label{eq:defofCpi44456} .
\end{align}
As a consequence of   Hypothesis \ref{H1} and the fact that $\psi_1$ and $\psi_2$ satisfy Hypothesis \ref{H2}, the function  on
the right hand side of  \eqref{eq:defofCpi44456} is integrable with respect to  the measure $du_0  \prod_{s \in I_A}  dv_s$ and the resulting integral
is bounded by 
\begin{align} 
\frac{ \| \widehat{B}(- e^{- i \vartheta \cdot } )\|_1^{|I_A|}
 \| \widehat{B}(- e^{- i \vartheta \cdot } )\|_\infty^{|A|} }{ |\sin(2 \vartheta ) E  |^{n+1}}   \left[  \int du_0   \left| \overline{ \widehat{\psi}}_1(e^{   i \vartheta}u_0)  \right|^2 \right]^{1/2}  \left[ \int du_0
\left| \widehat{\psi}_2 ( e^{ -  i \vartheta} u_0  )  \right|^2 \right]^{1/2}    \label{majofanaint} 
 . 
\end{align}
 The existence of the limit $\eta \downarrow 0$ of \eqref{eq:defofCpi222}
 thus follows by first fixing a $\vartheta \in (0,{\rm min}(\vartheta_{B},\vartheta_{\psi},\pi/4))$ and then using  the  dominated convergence theorem.
 Note that dominated convergence 
 is justified by the integrable majorant  \eqref{majofanaint} and the  continuity of the integrand of \eqref{eq:defofCpi222} in $\eta \geq 0$   for fixed $E > 0$. 
Analogously,  it follows that $T_{n,\infty}[z ; \psi_1 , \psi_2]$ has  a continuous 
extension to $\{ w \in \C : {\rm Im } w \geq  0, \ {\rm Re } w > 0 \}$ for $z \in \{ w \in \C : {\rm Im } w > 0, \ {\rm Re } w > 0 \}$,  by choosing the same majorant and using the continuity of the integrand in $z = E + i \eta$. 
This completes the proof Theorem  \ref{thm:boundaryvelueexpcoeff}.
\end{proof}

\section{Asymptotic Error Estimate} 
\label{sec:AsymEE}
The goal of this  section is to  prove Theorem \ref{thm:maintec000}, which will take some steps as preparatory work. Let $z \in \C \setminus [0,\infty)$. 
For $\lambda \geq 0$, we start of with recalling the   expansion \eqref{eq:resexp} 
\begin{align*} 
( H_L  - z)^{-1} & = \sum_{j=0}^{n-1} R_L(z) [ \lambda V_L R_L(z)  ]^j +  [R_L(z)  \lambda V_L  ]^{n}  ( H_L  - z)^{-1} . 
\end{align*}
Recalling $T_{j,L}$ defined in \eqref{defoffinT}, we can write the expectation of matrix elements  as  
\begin{align} \label{eq:expansionU} 
 {\bf E }_L \inn{ \psi_{1,\#} , ( H_L  - z )^{-1} \psi_{2,\#} }_{L^2(\Lambda_L)}   & = \sum_{j=0}^{n-1}  \lambda^j T_{j,L} [z ; \psi_1, \psi_2]  +  \lambda^{n} U_{n,L}[z ;  \psi_1 , \psi_2 ; \lambda]   , 
\end{align}
where we defined  
\begin{align}\label{eq:expansionU2} 
 U_{n,L}[z ;  \psi_1 , \psi_2 ; \lambda] &  :=  {\bf E }_L \inn{ \psi_{1,\#} ,  [R_L(z)   V_L  ]^{n}  ( H_L - z  )^{-1}  \psi_{2,\#} }_{L^2(\Lambda_L)}   .
\end{align}

In this section we control the error terms $U_{n,L}$.  
For the error estimate we use the Cauchy-Schwarz inequality twice and find 
\begin{align}
|   U_{n,L}[z ;  \psi_1 , \psi_2 ; \lambda]   | &={\bf E }_L \inn{ \left(  [R_L(z)   V_L  ]^{n}  \right)^* \psi_{1,\#} ,   ( H_L - z  )^{-1} 
 \psi_{2,\#} }_{L^2(\Lambda_L)}  \\
& \leq {\bf E}_L  \|    [(R_L(z)   V_L  )^*]^{n}   \psi_{1,\#} \|_{L^2(\Lambda_L)}   \|  ( H_L - z  )^{-1}  \psi_{2,\#}  \|_{L^2(\Lambda_L)}   \nonumber \\
& \leq   ( E_{n,L}[z ; \psi_1])^{1/2}    |{\rm Im} z |^{-1} \| \psi_{2,\#} \|_{L^2(\Lambda_L)}      \label{eq:boundonU}  , 
\end{align}
where we defined 
\begin{align} \label{defofENL} 
E_{n,L}[z ; \psi_1]   & :=   {\bf E}_L  \|    [(R_L(z)   V_L  )^*]^{n}   \psi_{1,\#} \|_{L^2(\Lambda_L)} ^2 \\
& =    {\bf E}_L  \inn{ \psi_{1,\#} ,  [(R_L(z)   V_L  )]^{n}   [V_L R_L(\overline{z}) ]^{n}   \psi_{1,\#} }_{L^2(\Lambda_L)}    \nn
\end{align}
and used the basic inequality \eqref{ln-1} together with the spectral theorem. 

To estimate  \eqref{defofENL}  transition to the Fourier space using identities  \eqref{ftident1} and \eqref{ftident2}, and  note that 
between the middle $V_L$'s we have instead of a  resolvent an  identity, which we write for bookkeeping  in terms of   a delta function as in  \eqref{deltaL} 
(note that Fubinis theorem is applicable since $z \in \C \setminus [0,\infty)$ - hence the resolvents are well defined - and the assumptions on $\hat{\psi}$ and $\widehat{V_L}$ and $\widehat{B}$ respectively by \eqref{secondRest})
\begin{align}
&    E_{n,L}[z ; \psi_1]  \nonumber \\ 
&  =  \int_{(\Lambda_L^*)^{n+1} }  \int_{(\Lambda_L^*)^{n+1} } \overline{  \widehat{\psi}}_{1,\#}(p_1) \widehat{\psi}_{1,\#}(\tilde{p}_{n+1})  {\bf E}_L   \prod_{j=1}^n  \widehat{V}_{L}(p_j - p_{j+1} ) \delta_{*,L}(p_{n+1} - \tilde{p}_1)  \prod_{j=1}^n  \widehat{V}_{L}(\tilde{p}_j - \tilde{p}_{j+1} )  
\nonumber  \\
& \times  \left(    \prod_{j=1}^{n}      \frac{1}{ \nu(p_j) - z}        \right)  \left(    \prod_{j=2}^{n+1}      \frac{1}{ \nu(\tilde{p}_j) - \overline{z}}        \right) d(p_1, \ldots , p_{n+1})  d(\tilde{p}_1, \ldots ,  \tilde{p}_{n+1}) 
 \nonumber \\
&  = \int_{(\Lambda_L^*)^{2n+1} }  \overline{   \widehat{\psi}}_{1,\#}(q_1) \widehat{\psi}_{2,\#}(q_{2n+1})   {\bf E}_L   \prod_{j=1}^{2n}  \widehat{V}_{L}(q_j - q_{j+1} )
\nonumber  \\
& \times  \left(    \prod_{j=1}^{n}      \frac{1}{ \nu(q_j) - z}        \right)  \left(    \prod_{j=n+2}^{2n+1}      \frac{1}{ \nu(q_j) - \overline{z}}        \right) d (q_1 , \ldots , q_{2n+1}). \label{eq:expforTn} 
\end{align} 
In the second equality we integrated out $p_{n+1}$ using the delta function and introduced the following simple relabeling of the integration variables 
$$
q=((q_1,...q_n),(q_{n+1},...,q_{2n+1})) = ((p_1,...,p_{n}),(\tilde{p}_1,...,\tilde{p}_{n+1})) =(p,\tilde{p}) .
$$
Now let us calculate the expectation in  \eqref{eq:expforTn}. 
Using    \eqref{eq:ofexpofrandpot2}  in  \eqref{eq:expforTn}, we find 
\begin{align}
   E_{n,L}[z ; \psi_1]  \label{EREL}  
& =  \int_{(\Lambda_L^*)^{2n+1}}  \sum_{A  \in \mathcal{A}_{2n}} \mathcal{P}_{A,L}( (q_1 , \ldots , q_{2n+1}))  \overline{ \widehat{  \psi}}_{1,\#}(q_1)     \widehat{\psi}_{1,\#}(q_{2n+1})
\\
&\times \prod_{j=1}^{n}  (\nu( q_j) - z)^{-1} \prod_{j=n+2}^{2n+1}  (\nu( q_j) - \overline{z})^{-1} d (q_1 , \ldots , q_{2n+1}),
\end{align}
where we  recall the definition given in 
 \eqref{eq:defofdeltapi0}
\begin{align} \label{eq:defofdeltapi}
\mathcal{P}_{A,L}(q) & =  \prod_{a \in A} \left[ m_{|a|}   \delta_{*,L} \left( \sum_{l \in a} (q_l - q_{l+1}) \right) 
  \prod_{l \in a}  \widehat{B}_\#(q_l - q_{l+1} ) \right]  .
 \end{align}
  For $A  \in \mathcal{A}_{2n} $ we define
\begin{align} \label{eq:defofCpitilde}
 D_{n,A,L}[z;\psi_1] 
& :=      \int_{(\Lambda_L^*)^{2n+1}}  \mathcal{P}_{A,L}( (q_1 , \ldots , q_{2n+1})) \overline{  \widehat{ \psi}}_{1,\#}(q_1)     \widehat{\psi}_{1,\#}(q_{2n+1}) \nn
\\
& \times\prod_{j=1}^{n}  (\nu( q_j) - z)^{-1}\prod_{j=n+2}^{2n+1}  (\nu( q_j) - \overline{z})^{-1} d (q_1 , \ldots , q_{2n+1}) 
\end{align}
With this definition we can    write    \eqref{EREL} as 
\begin{align} \label{CtildeErel} 
E_{n,L}[ \cdot  ; \cdot  ]  & = \sum_{A \in \mathcal{A}_{2n}}  D_{n, A,L}[ \cdot ; \cdot]  .  
\end{align}

To estimate \eqref{eq:defofCpitilde} 
we  introduce  the following change of variables  similar to   \eqref{changevar1}
\begin{align} \label{changevar1v2}
u_0  = q_1  \text{ and }  u_{l} = q_{l+1} - q_l \text{ for }  l=1,...,n 
\end{align}
and $u=(u_0 , \ldots ,  u_{2n})$.
 With this we can express the $q_j$'s as
$$
q_j = \sum_{l=0}^{j-1} u_l .
$$
Using the change of variables  \eqref{changevar1v2} in  \eqref{eq:defofCpitilde}  we find with \eqref{eq:defofdeltapi} expressed in the new variables 
that 
  for $A  \in \mathcal{A}_{2n} $ 
\begin{align}  \label{eq:defofCpiVarChange0}
D_{n,A,L}[z ; \psi_1] & =      \int_{(\Lambda_L^*)^{2n+1}}  \prod_{a \in A} \left[ m_{|a|}  \delta_{*,L} \left( \sum_{l \in a} u_l\right) 
  \prod_{l \in a}  \hat{B}_\#(- u_l) \right]  \overline{\widehat{ \psi}}_{1,\#}(u_0)     \widehat{\psi}_{1,\#} \left( \sum_{l=0}^{2n} u_l \right) \nn
\\
& \times    \prod_{j=1}^{n}  \left(\nu\left(  \sum_{l=0}^{j-1} u_l \right) - z \right)^{-1}\prod_{j=n+2}^{2n+1}  \left(\nu \left(  \sum_{l=0}^{j-1} u_l \right) - \overline{z} \right)^{-1}  d(u_0 , \ldots , u_{2_n}).     
\end{align}

We are now going to resolve the $\delta_{*,L} \left( \sum_{l \in a} u_l \right) $ factors similarly to the procedure presented between  \eqref{eq:defofCpi} and Lemma \ref{totsumzero}. We are going to use the notation  as introduced right before \eqref{changevar1next},  where for $A \in \mathcal{A}_{2n}$ we define
 \begin{align}  \label{defofJA} 
 J_A := \{ \max a : a \in A  \} \subset \{ 1 , \ldots 2n \} ,
\end{align}   as the set of all the largest elements of the respective sets $a \in A$ of the partition, as well as its complement
 \begin{align} \label{defofIA} 
I_A := \{ 1,...,2n \} \setminus J_A.
\end{align}  
For $l \in \{ 1, \ldots ,  2n \}$ we define $ a(l) \in A$ as the set with $l \in a(l)$.  Note that this choice is both, always possible as well as unique, since $A$ is a partition of  $\{ 1, \ldots ,  2n \}$.
With this we define the map 
\begin{align} \label{defofMA} 
M_A : \left( \R^d \right)^{ |I_A|} \to \left( \R^d \right)^{ 2n} \quad \text{via} \quad
[M_A (u)]_j =
\begin{cases}
u_j &: j \in I_A,
\\
- \sum\limits_{ l \in a (j)  \setminus \{ j \} }  u_l &: j \in  J_A.
\end{cases}
\end{align}

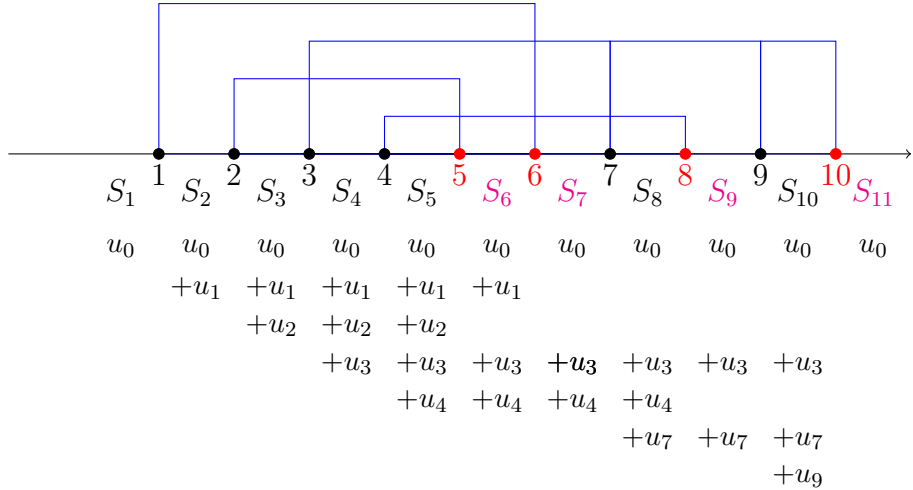
\begin{figure}[H]
\begin{tikzpicture}
\draw [draw=blue] (-3,0) rectangle (2,2);
\draw [draw=blue] (-2,0) rectangle (1,1);
\draw [draw=blue] (-1,0) rectangle (3,1.5);
\draw [draw=blue] (3,0) rectangle (5,1.5);
\draw [draw=blue] (5,0) rectangle (6,1.5);
\draw [draw=blue] (0,0) rectangle (4,0.5);
  \draw[->] (-5, 0) -- (7, 0) ;
\filldraw[black]  (-3,0)  circle[radius=2pt] node[below]{1};
\filldraw[black]  (-2,0)  circle[radius=2pt]  node[below]{2}; ;
\filldraw[black]  (-1,0)  circle[radius=2pt] node[below]{3};  ;
\filldraw[black]  (0,0)  circle[radius=2pt]  node[below]{4};  ;
\filldraw[red]  (1,0)  circle[radius=2pt] node[below]{5}; ;
\filldraw[red]  (2,0)  circle[radius=2pt] node[below]{6};  ;
\filldraw[black]  (3,0)  circle[radius=2pt] node[below]{7};;
\filldraw[red]  (4,0)  circle[radius=2pt] node[below]{8}; ;
\filldraw[black]  (5,0)  circle[radius=2pt]  node[below]{9}; ;
\filldraw[red]  (6,0)  circle[radius=2pt] node[below]{10};   ;
\draw  (-3.5,-0.2)   node[below] {$S_1$};
\draw  (-2.5, -0.2 )   node[below] {$S_2$};
\draw  (-1.5, -0.2  )   node[below] {$S_3$};
\draw  (-0.5,-0.2 )   node[below] {$S_4$};
\draw  (0.5, -0.2 )   node[below] {$S_5$};
\draw[magenta]  (1.5, -0.2 )   node[below] {$S_6$};
\draw[magenta]  (2.5, -0.2 )   node[below] {$S_7$};
\draw  (3.5, -0.2 )   node[below] {$S_8$};
\draw[magenta]  (4.5,-0.2  )   node[below] {$S_9$};
\draw  (5.5, -0.2  )   node[below] {$S_{10}$};
\draw[magenta]  (6.5,  -0.2)   node[below] {$S_{11}$};
\draw  (-3.5,-1)   node[below] {\small $u_0$ };
\draw  (-2.5,-1)   node[below] {\small $u_0$};
\draw  (-2.5,-1.5)   node[below] {\small $+u_1$};
\draw  (-1.5,-1)   node[below] {\small $u_0$};
\draw  (-1.5,-1.5)   node[below] {\small $+u_1$};
\draw  (-1.5,-2)   node[below] {\small $+u_2$};
\draw  (-0.5,-1)   node[below] {\small $u_0$};
\draw  (-0.5,-1.5)   node[below] {\small $+u_1$};
\draw  (-0.5,-2)   node[below] {\small $+u_2$};
\draw  (-0.5,-2.5)   node[below] {\small $+u_3$};
\draw  (0.5,-1)   node[below] {\small $u_0$};
\draw  (0.5,-1.5)   node[below] {\small $+u_1$};
\draw  (0.5,-2)   node[below] {\small $+u_2$};
\draw  (0.5,-2.5)   node[below] {\small $+u_3$};
\draw  (0.5,-3)   node[below] {\small $+u_4$};
\draw  (1.5,-1)   node[below] {\small $u_0$};
\draw  (1.5,-1.5)   node[below] {\small $+u_1$};
\draw  (1.5,-2.5)   node[below] {\small $+u_3$};
\draw  (1.5,-3)   node[below] {\small $+u_4$};
\draw  (2.5,-1)   node[below] {\small $u_0$};
\draw  (2.5,-2.5)   node[below] {\small $+u_3$};
\draw  (2.5,-3)   node[below] {\small $+u_4$};
\draw  (3.5,-1)   node[below] {\small $u_0$};
\draw  (3.5,-3)   node[below] {\small $+u_4$};
\draw  (4.5,-1)   node[below] {\small $u_0$};
\draw  (2.5,-2.5)   node[below] {\small $+u_3$};
\draw  (3.5,-2.5)   node[below] {\small $+u_3$};
\draw  (4.5,-2.5)   node[below] {\small $+u_3$};
\draw  (5.5,-2.5)   node[below] {\small $+u_3$};
\draw  (5.5,-1)   node[below] {\small $u_0$};
\draw  (6.5,-1)   node[below] {\small $u_0$};
\draw  (3.5,-3.5)   node[below] {\small $+u_7$};
\draw  (4.5,-3.5)   node[below] {\small $+u_7$};
\draw  (5.5,-3.5)   node[below] {\small $+u_7$};
\draw  (5.5,-4)   node[below] {\small $+u_9$};
\end{tikzpicture}

\caption{  \label{bild2} The blue lines connect the points which belong to the same set $a$ of the partition $A$.  The elements of the set $J_A$ are coloured with red and the elements of the set $I_A $ are black.
The $S_j$ such that $j-1\in J_A$ are coloured with magenta. 
}
\end{figure}

In the following we will be working with terms of the form  $S_j = u_0 + \sum\limits_{l=1}^{j-1}[M_A(u)]_l$.  To get some intuition on the defined objects let us discuss the following example and  illustrate it with a picture: We consider the  set $\{1,...,10\}$ with  $n=5$ together with the partition $A = \{ \{1,6\} ,\{2,5\},\{3,7,9,10\}, \{4,8\} \}$ .  Therefore $J_A =\{5,6,8,10\}$ and $I_A =\{1,2,3,4,7,9\}$.  The picture displays $S_1$ to $S_{11}$.
This is illustrated in the Figure \ref{bild2}.

Using this notation to express   \eqref{eq:defofCpiVarChange0}
 we find  for $A  \in \mathcal{A}_{2n} $ 
\begin{align} \label{eq:defofCpiVarChange}
& D_{n,A,L}[z ;\psi_1] \\
&=      \int_{\Lambda_L^*} d u_0  \prod_{l \in I_A} \left( \int_{\Lambda_L^*} d u_l  \right)   \prod_{a \in A} \left[ m_{|a|}   \prod_{l \in a }  \hat{B}_\#(- [M_A (u)]_l) \right] \overline{\widehat{\psi}}_{1,\#}(u_0)     \widehat{\psi}_{1,\#} \left( u_0  \right) \nn
\\
& \times \prod_{j=1}^{n}  \left(\nu\left( u_0 + \sum_{l=1}^{j-1} [M_A (u)]_l  \right) - z  \right)^{-1}\prod_{j=n+2}^{2n+1}  \left (\nu \left( u_0 + \sum_{l=1}^{j-1} [M_A (u)]_l  \right) - \overline{z}  \right)^{-1}, \nn
\end{align}
where we used  $\sum_{j=1}^{2n}  [M_A (v)]_j = 0$ by Lemma \ref{sum0} in the argument of the second $ \widehat{\psi}_{1,\#} $.

We are now in a situation in which resolving the $\delta$-functions eliminated the  sums over the $u_j$ for $j-1 \in J_A$,  while the sums for $j-1 \in I_A \cup \{0 \}$ remain.  The idea is now that each ``integral'' $d u_j$ can control with  the respective $\hat{B}_\#( -u_j)$  one resolvent,  which we will show in Lemma \ref{ausintegrieren}.  We will first estimate the other resolvents as well as the $\hat{B}_\#( -u_j)$ that belong to $j \in J_A$.
 
We write  $\eta={\rm Im} (z)$.  We estimate  all  resolvents for which $j-1 \in J_A$ by a factor of $\eta^{-1}$ using  \eqref{ln-1},  assuming $0 < \eta \leq 1$.  
Note that this may include  the case where  $j = n+1$ and $j-1 \in J_A$, that is 
where we have an identity instead of a  resolvent, in which case  we    estimate trivially by $\eta^{-1}$. 
Since $J_A$ contains one element of each partition set in $A$,  we have $| J_A | = |A|$ and therefore pay a factor of  $\eta^{-|A|}$.
In addition, to obtain an $L^2$--bound for the wave function $\psi_1$,  for convenience, we estimate the first  resolvent, i.e.,  $j=1$, with $\eta^{-1}$, 
and therefore pay a total factor of  $\eta^{-|A|-1}$, since $1 \notin J_A$ by the assumption $m_1=0$. 
Furthermore,   for $j \in J_A$ we estimate  $|\hat{B}_\#(-[M_A(u)]_j)| \leq  \| \hat{B}_\#\|_{*,\infty} $ trivially, which gives a collective 
upper bound  for these terms by  $ \| \hat{B}_\#\|_{*,\infty}^{|A|}$, again since $| J_A |=| A |$.  After these  steps we arrive at 
\begin{align}  \label{Dtoestim} 
&| D_{n,A,L}[z ;\psi_1] | \\
& \leq   \eta^{-|A|-1}  \| \hat{B}_\# \|_{*,\infty}^{|A|} \left|  \prod_{a \in A}  m_{|a|}  \right|  \int_{\Lambda_L^*} d u_0   \prod_{l \in I_A}  \left( \int_{\Lambda_L^*} d u_l  \right)   \left|  \widehat{\psi}_{1,\#}(u_0)  \right|^2  \bigg|    \prod_{l=1, l \in I_A }^{2n}   \hat{B}_\#( -u_l) \ \bigg|
 \nonumber  \\
& \times \prod_{\substack{ j=2 \\ j - 1 \in I_A}}^{n}  \left|\nu\left( u_0 + \sum_{l=1}^{j-1}[M_A (u)]_l  \right) - z  \right|^{-1}\prod_{\substack{ j=n+2 \\  j - 1 \in I_A}}^{2n+1}  \left| \nu \left( u_0 + \sum_{l=1}^{j-1} [M_A (u)]_l   \right) - \overline{z}  \right|^{-1} \nonumber .  
\end{align}

To estimate  the remaining resolvents, we shall use the Lemma  \ref{lemmalinindep}, below. We view $u_j$  as the function
\begin{align}
\label{uj}
u_j : \R^{2n+1} \to \R,  \quad(t_0,...,t_{2n}) \mapsto t_j \text{ for } j=0, \ldots, 2n .
\end{align}
In the following lemma,  we will show that the functions  $\{ u_0 + \sum_{l=1}^{j-1}[M_A (u)]_l : j -1 \in I_A \} $   are linearly independent.  Doing so we will calculate the sums explicitly and show
that  for all $j=1,...,2n+1$ with $j-1 \in I_A$ we have a relation   of the form
$$
 u_0 + \sum_{l=1}^{j-1}[M_A (u)]_l   = u_{j-1} + u_0  + \text{a linear combination of } \{ u_l : l \leq j-2, l \in I_A\} .
$$
The identity above will later be used in Lemma \ref{ausintegrieren} to integrate over the $d u_l$ backwards (beginning with the one with the largest  index) one after another to eliminate the respective $u_l$.

\begin{lemma} \label{lemmalinindep} 
Let $A$ be a partition  of $\{ 1, \ldots ,  2n \}$.  Let $J_A,I_A$ and $M_A$ be defined as in \eqref{defofJA},   \eqref{defofIA} and  \eqref{defofMA}.  Let $u_j$ for $j \in \{ 0, \ldots ,  2n \}$ be the function in \eqref{uj}.
Let for $A$ and $j =1,...,2n+1$ the function  $\sigma_{A,j}$ be defined as
\begin{align*}
\sigma_{A,j} : \{ 1 ,  \ldots ,  2n \} \to \{ 0,1 \} \quad \text{via} \quad
\sigma_{A,j}  (l) =
 \begin{cases}
1 \quad : \quad  \max {a(l)} > j,
\\
0 \quad : \quad \text{otherwise}.
\end{cases}
\end{align*}

Then for all $j=2,...,2n+1$ with $j-1 \in I_A$ we have
\begin{align} \label{linindep} 
 u_0 + \sum_{l=1}^{j-1}[M_A (u)]_l   = u_{j-1}  + u_0 + \sum_{l \in \{1,...,j-2\} \cap    I_A } \sigma_{A , j-2}  (l) u_l .
\end{align}
\end{lemma}

\begin{proof}
To shorten the notation let $N_{k} = \{1,...,k\}$ for $k \in \N$.
Plugging  in the definition of  $M_A$  for its respective cases we get    for $j -1 \in I_A$ 
\begin{align*}
 u_0 + \sum_{l=1}^{j-1}[M_A (u)]_l  
 &=    u_0 + \sum_{l \in N_{j-1} \cap   I_A } [M_A (u)]_l  +  \sum_{l \in N_{j-1} \cap   J_A }[M_A (u)]_l  
 \\
 & =u_0 + \sum_{l \in N_{j-1} \cap   I_A } u_l  +  \sum_{l \in N_{j-1} \cap   J_A } \sum_{ s \in a (l)  \setminus \{ l \} } ( -  u_s)
\\
 & \overset{j -1 \notin J_A}{=}u_0 + u_{j-1} + \sum_{l \in N_{j-2} \cap   I_A } u_l   + \sum_{l \in N_{j-2} \cap   J_A }  \sum_{ s \in a (l)  \setminus \{ l \} } (-  u_s )
\\
 & \overset{(*)}{=}u_0 + u_{j-1} + \sum_{l \in N_{j-2} \cap   I_A } u_l   + \sum_{ \substack{l \in N_{j-2} \cap   I_A \\ \max{a(l)} \leq j-2}} ( - u_l ) 
\\
&= u_{j-1}  + u_0 + \sum_{ \substack{l \in N_{j-2} \cap   I_A \\ \max{a(l)} > j-2}} u_l   + \sum_{ \substack{l \in N_{j-2} \cap   I_A \\ \max{a(l)} \leq j-2}} u_l  + 
\sum_{ \substack{l \in N_{j-2} \cap   I_A \\ \max{a(l)} \leq j-2}}  (- u_l ) 
 \\
 &= u_{j-1}  + u_0 + \sum_{ \substack{l \in N_{j-2} \cap   I_A \\ \max{a(l)} >j-2}} u_l  + \sum_{ \substack{l \in N_{j-2} \cap   I_A \\ \max{a(l)} \leq j-2}} ( u_l - u_l )
 \\
 &= u_{j-1}  + u_0 + \sum_{l \in N_{j-2} \cap   I_A }  \sigma_{A , j-2}  (l) u_l .
\end{align*}
Now to $(*)$: We have to show
\begin{align}\label{eq:setequality-1} 
\sum_{l \in N_{j-2} \cap   J_A }  \sum_{ s \in a (l)  \setminus \{ l \} } -  u_s =   \sum_{ \substack{l \in N_{j-2} \cap   I_A \\ \max{a(l)} \leq j-2}} - u_l.
\end{align}
From the proof  of Lemma \ref{sum0},  we have seen
\begin{align} \label{eq:setequality0} 
I_A = \bigcup_{l \in J_A} a(l) \setminus \{ l \} . 
\end{align}
Using this we will now show  that 
\begin{align} \label{eq:setequality} 
I_A \cap \{ l : \max a(l) \leq j-2 \}  &  =   \bigcup_{l \in N_{j-2} \cap  J_A} a(l) \setminus \{ l \} 
\end{align}
First $\subset$: Let $s$ be an element of the LHS of \eqref{eq:setequality}. Since $s \in I_A$, we have $s \in a(l) \setminus \{ l \}$ for some $l \in J_A$ by 
\eqref{eq:setequality0}. Now since s in $\{ t : \max a(t) \leq j-2 \}$  we find  $j-2 \geq \max a(s) = \max(a(l)) = l$.  \\
 Second   $\supset$:  Let $\mathcal{L}$ denote the set on the RHS of \eqref{eq:setequality}. Then $\mathcal{L} \subset I_A$ (by  \eqref{eq:setequality0}).
 If $k \in L$, then $k \in a(l) \setminus \{ l \}$ for some $l \in N_{j-2} \cap J_A$
and so $\max a(k) =\max a(l) =  l \leq j-2$. \\

Moreover, we have 
\begin{align} \label{eq:setequality2} 
I_A \cap \{ l : \max a(l) \leq j-2 \}   \subset  N_{j-3} 
\end{align}
since for $s \in I_A$ with $\max a(s) \leq j-2$,  we find  $s + 1   \leq  \max a(s) \leq j-2$. 

   Now \eqref{eq:setequality-1}  follows from \eqref{eq:setequality},  \eqref{eq:setequality2},  and the fact that the elements $a \in A$ are mutually disjoint. 

\end{proof}
As a final step before we can prove Theorem \ref{thm:maintec000},  we will use the expression shown in Lemma \ref{lemmalinindep}, performing an induction to integrate over the $u_l$ variables one after another.
\begin{lemma}
\label{ausintegrieren}
Let $n \in \N$ and $A$ be a partition of the set $\{ 1, \ldots ,  2n \}$.  Let $J_A,I_A$ and $M_A$ be defined as in \eqref{defofJA},   \eqref{defofIA} and  \eqref{defofMA}.  Then  for $z \in \{ w \in \C : {\rm Im } \, w > 0 \}$ we have 
\begin{align}   \label{LHS}
&\int_{\Lambda_L^*} d u_0   \prod_{l \in I_A}  \left( \int_{\Lambda_L^*} d u_l  \right)   | \overline{ \hat{\psi}}_{1,\#}(u_0)  |^2  \bigg|    \prod_{l=1, l \in I_A }^{2n}  \hat{B}_\#( -u_l)  \bigg|
 \\
 &\times
 \prod_{\substack{ j=2 \\ j - 1 \in I_A }}^{n}  \left|\nu\left( u_{j-1}  + u_0 + \sum_{l \in \{1,...,j-2\} \cap    I_A } \sigma_{A , j-2}  (l) u_l \right) - z  \right|^{-1}  \nonumber   \\
  &\times
 \prod_{\substack{ j=n+2 \\ j - 1 \in I_A }}^{2n+1}  \left|\nu\left( u_{j-1}  + u_0 + \sum_{l \in \{1,...,j-2\} \cap    I_A } \sigma_{A , j-2}  (l) u_l \right) - \overline{z} \right|^{-1}
 \nonumber \\
 &\leq \| \psi_1 \|^2  [  {C}(E,d,L,\eta,\hat{B}_\#)   ]^{2n-|A|}  , \nonumber 
\end{align}
where $\sigma_{A, j }$ is defined in Lemma   \ref{lemmalinindep}  and  ${C}(E,d,L,\eta,\hat{B}_\#)$ is the constant \eqref{C007} from Proposition \ref{ln000}.
\end{lemma}

\begin{proof}
Let  $z = E +   i \eta$ with $E  > 0$ and $\eta > 0$ (the case $E-i \eta$ then follows by taking complex conjugates).  
Moreover, since $|a+ib|= (a^2 + b^2 )^{1/2}$ for $a,b \in \R$ and hence does not depend on the sign of $b$,
we can replace $\overline{z}$ with $z$ in the above formula without change. 
To shorten the writing expenses we introduce some notation: Let  $P := P(E,d,L,\eta,\hat{B}_\#, \hat{\psi})  $ denote  the LHS  of \eqref{LHS},
 i.e., 
\begin{align*}  
P=&\int_{\Lambda_L^*} d u_0   \prod_{l \in I_A}  \left( \int_{\Lambda_L^*} d u_l  \right)   | \overline{ \hat{\psi}}_{1,\#}(u_0)  |^2  \bigg|    \prod_{l=1, l \in I_A }^{2n} \hat{B}_\#( -u_l)  \bigg|
 \nonumber  \\
 &\times
 \prod_{\substack{ j=2 \\ j - 1 \in I_A \\ j \neq n + 1}}^{2n+1}  \left|\nu\left( u_{j-1}  + u_0 + \sum_{l \in \{1,...,j-2\} \cap    I_A } \sigma_{A , j-1}  (l) u_l \right) - E - i \eta \right|^{-1}.  \nonumber
\end{align*}
Since we now want to integrate out the variables one after another,  starting with the one with the largest index,  we consider the set  $I_A$, but excluding it's $m$ largest elements,  which we call $K_m$.  So, in that sense let $K_0=I_A$ and $K_1 = I_A \setminus \{ \max I_A \}$ and so on, explicitly  $K_m = \{ j \in I_A :  | \{ s \in I_A : s > j \} |  \geq  m \}$. 

We are now going to prove the following statement in $m \in \{ 0, \ldots , 2n+1  \}$ with an induction over $m$ [Note that we shift the index $j$  by $1$, i.e., we  write $j$ instead of $j-1$.]
\begin{align}  
P&\leq  [    C(E,d,L,\eta,\hat{B}_\#)   ]^m 
 \int_{\Lambda_L^*} d u_0   \prod_{ l \in K_m}  \left( \int_{\Lambda_L^*} d u_l  \right)   | \overline{ \hat{\psi}}_{1,\#}(u_0)  |^2  \bigg|    \prod_{l \in K_m}   \hat{B}_\#( -u_l)  \bigg|
 \nonumber  \\
 &\times
 \prod_{\substack{ j \in K_m \\ j  \neq n }} \left|\nu\left( u_{j}  + u_0 + \sum_{l \in \{1,...,j-1\} \cap    I_A } \sigma_{A , j}  (l) u_l \right) - E - i  \eta \right|^{-1}  \nonumber   .     \nonumber 
\end{align}

Since we can not take more elements out of the set $I_A$ other than those that are in $I_A$,  the variable $m$ can not exceed	the value $| I_A  | = 2n-  | A |$. Therefore for $0 \leq m \leq 2n - | A |$ this statements makes sense. 

First consider  $m=0$.  In this case $[  C(E,d,L,\eta,\hat{B}_\#)   ]^m=1$ and $K_m=K_0=I_A$ hence the RHS is equal to the definition of $P$. 
Now let the statement hold true for  an $m$ with $0 \leq m \leq 2n - | A | -1$.  Let $s= \max K_m$.
Our plan is to use Proposition \ref{ln00} to eliminate the ``integral'' over $u_s$.  Now only $\hat{B}_\#( -u_{s})$ and, by Lemma \ref{linindep},  the  resolvent with  $j =
s $ depend on $u_s$.

To create the proper form for Proposition \ref{ln00} to be applicable,  we perform a change of variables given by
$
v= u_{s}  + u_0 + \sum_{l \in \{1,...,s-1\} \cap    I_A } \sigma_{A , s}  (l) u_l 
$
and therefore
\begin{align*}
-u_s = - v + u_0 + \sum_{l \in \{1,...,s-1\} \cap    I_A } \sigma_{A , s}  (l) u_l.
\end{align*}
We also introduce the notation
\begin{align*}
\tilde{B}(v):= \hat{B}_\# \left( - v + u_0 + \sum_{l \in \{1,...,s-1\} \cap    I_A } \sigma_{A , s}  (l) u_l \right).
\end{align*}
We are now able to write
\begin{align*}
& \int_{\Lambda_L^*}  | \hat{B}_\#( -u_{s})  | \left|\nu\left( u_{s}  + u_0 + \sum_{l \in \{1,...,s-1 \} \cap    I_A } \sigma_{A , s}  (l) u_l \right) - E - i \eta \right|^{-1}  d u_{s}
\\&=\int_{\Lambda_L^*} | \tilde{B}(v) |  \left|\nu\left(v \right) - E - i \eta \right|^{-1}  d v \\
&\leq  C(E,d,L,\eta,\tilde{B}) = {C}(E,d,L,\eta,\hat{B}_\#)   ,
\end{align*}
where we applied Proposition \ref{ln000} in the inequality.  Note that  ${C}(\cdots)$  being the constant given in Proposition \ref{ln000}, where the definition \eqref{C007} of $C$ gives the final equality.
Plugging this into the statement for $m$, recall  $s=\max K_m$,  gives us
\begin{align}  
P &\leq  [   {C}(E,d,L,\eta,\hat{B}_\#)   ]^m 
 \int_{\Lambda_L^*} d u_0   \prod_{ l \in K_m}  \left( \int_{\Lambda_L^*} d u_l  \right)   | \overline{ \hat{\psi}}_{1,\#}(u_0)  |^2  \bigg|    \prod_{l \in K_m}  \hat{B}_\#( -u_l)  \bigg|
 \nonumber  \\
 &\times
 \prod_{\substack{ j \in K_m \\ j \neq n }} \left|\nu\left( u_{j}  + u_0 + \sum_{l \in \{1,...,j-1\} \cap    I_A } \sigma_{A , j}  (l) u_l \right) - E - i \eta \right|^{-1}  \nonumber    \nonumber \\
&= [  {C}(E,d,L,\eta,\hat{B}_\#)   ]^m 
  \int_{\Lambda_L^*} d u_0   \prod_{ l \in K_{m+1}}  \left( \int_{\Lambda_L^*} d u_l  \right)   | \overline{ \hat{\psi}}_{1,\#}(u_0)  |^2  \bigg|    \prod_{l \in K_{m+1}}   \hat{B}_\#( -u_l)  \bigg|
 \nonumber  \\
 &\times
 \prod_{\substack{ j \in K_{m+1} \\ j \neq n }} \left|\nu\left( u_{j}  + u_0 + \sum_{l \in \{1,...,j-1\} \cap    I_A } \sigma_{A , j}  (l) u_l \right) - E - i \eta \right|^{-1}  \nonumber    \nonumber \\
 &\times \int_{\Lambda_L^*} | \hat{B}_\#( -u_{s}) |   \left|\nu\left( u_{s}  + u_0 + \sum_{l \in \{1,...,s-1 \} \cap    I_A } \sigma_{A , s}  (l) u_l \right) - E - i \eta \right|^{-1}  d u_{s} \nonumber \\
 &\leq  [  {C}(E,d,L,\eta,\hat{B}_\#)   ]^{m+1} 
 \int_{\Lambda_L^*} d u_0   \prod_{ l \in K_{m+1}}  \left( \int_{\Lambda_L^*} d u_l  \right)   | \overline{ \hat{\psi}}_{1,\#}(u_0)  |^2  \bigg|    \prod_{l \in K_{m+1}} \hat{B}_\#( -u_l)  \bigg|
 \nonumber  \\
 &\times
 \prod_{\substack{ j \in K_{m+1} \\ j\neq n }} \left|\nu\left( u_{j}  + u_0 + \sum_{l \in \{1,...,j-1\} \cap    I_A } \sigma_{A , j}  (l) u_l \right) - E - i  \eta \right|^{-1}  .  \nn
\end{align}
This completes the induction over $m$.

Finally,  let  $m=2n-|A| = |I_A|$. Using   $K_{|I_A|} =\emptyset$,   the convention $\prod_{j \in \emptyset } a_j =1$, as well as 
  $\int_{\Lambda^*_L}|\hat{\psi}_{1,\#}(u_0)|^2 d u_0 = \|\psi_{1}\|_{L^2(\Lambda_L)}^2 \leq \|\psi_1\|^2$, by  Lemma  \ref{fourierseries} (a),  we  find
\begin{align}  
P&\leq  [   {C} (E,d,L,\eta,\hat{B}_\#)   ]^{2n-|A| }
 \int_{\Lambda_L^*} d u_0   \prod_{ l \in \emptyset}  \left( \int_{\Lambda_L^*} d u_l  \right)   | \overline{ \hat{\psi}}_{1,\#}(u_0)  |^2  \bigg|    \prod_{l \in \emptyset}   \hat{B}_\#( -u_l) \bigg|
 \nonumber  \\
 &\times
 \prod_{\substack{ j \in \emptyset \\ j +1 \neq n }} \left|\nu\left( u_{j}  + u_0 + \sum_{l \in \{1,...,j-1\} \cap    I_A } \sigma_{A , j}  (l) u_l \right) - E - i \xi_n (j) \eta \right|^{-1}  \nonumber    \nonumber \\
 &=  [    {C}(E,d,L,\eta,\hat{B}_\#)   ]^{2n-|A| }
 \int_{\Lambda_L^*}  | \overline{ \hat{\psi}}_{1,\#}(u_0)  |^2  d u_0   \nonumber   \\
 &\leq \|\psi_1\|^2 [  {C}(E,d,L,\eta,\hat{B}_\#)   ]^{2n-|A| } , \nonumber 
\end{align}
which completes the proof.
\end{proof}

We are now able to prove Theorem \ref{thm:maintec000} as outlined below.

\begin{proof}[Proof of Theorem \ref{thm:maintec000}]  We assume that $z = E + i \eta$ (the case $z = E - i \eta$ is obtained by taking complex conjugates). 
We start from \eqref{Dtoestim}. Inserting  \eqref{linindep} of Lemma  \ref{lemmalinindep} into  \eqref{Dtoestim} and  then using  Lemma  \ref{ausintegrieren} we find 

\begin{align} \label{0finalestonD}
|   D_{n,A,L}[E + i \eta ;\psi_1]|  & \leq    \eta^{-|A|-1}   \| \psi_1 \|^2 \| \hat{B}_\# \|_{*,\infty}^{|A|}    C(E,d, L, \eta,\hat{B}_\#)^{2n-|A|}   \left|  \prod_{a \in A} m_{|a|} \right|,
\end{align}
where $C(E,d, L, \eta,\hat{B}_\#) $ is by  Lemma \ref{ausintegrieren} the constant  \eqref{C007} given in Proposition \ref{ln000}
\begin{align} \label{CdEB0}
  &C(E,d, L, \eta,\hat{B}_\#) \\
  &=
2 \| \hat{B}_\#  \|_{*,\infty}  \left[  C_{1}(2 E,d)  \ln \left|  \eta^{-1}  +1    \right| +  2^d \sqrt{2} (4 E + 1)^{d/2} \right] + 2  \| \hat{B}_\# \|_{*,1}  .  \nn
\end{align} 

To estimate the $\| \widehat{B}_\# \|_{*,\infty}$    term appearing in \eqref{CdEB0}  we use  that by Lemma  \ref{fourierseries} (b) 
we have 
\begin{align*}
& \| \widehat{B}_{\#} \|_{*,\infty}  \leq \| B \|_1 < \infty  .
\end{align*}
Finally let us define $c_B$ as the RHS of  \eqref{secondRest}, which is an  $L$-independent upper bound on the $\| \hat{B}_\# \|_{*,1}$ term  appearing  in
 \eqref{CdEB0}.  
Now by \eqref{C007} this  shows that  \eqref{CdEB0}   is uniformly bounded in $L$.
Therefore we have
\begin{align} 
 & C(E,d, L, \eta,\hat{B}_\#) \nn \\
&\leq 2 \| B \|_1  C_{1}(2 E,d)  \ln \left|  \eta^{-1}  +1    \right| +  \left[ 2 \| B \|_12^d \sqrt{2} (4 E + 1)^{d/2}+ 2  c_B \right] \nn
\\
&\leq \tilde{C}(d,E,B) \left( \ln \left|  \eta^{-1}  +1    \right|  +1 \right), \label{CdEB}
\end{align} 
for \begin{align} \label{defoftildeCexplicit}  \tilde{C}(d,E,B):= \max \{ 2 \| B \|_1  C_{1}(2 E,d),  2 \| B \|_12^d \sqrt{2} (4 E + 1)^{d/2}+ 2  c_B \} . \end{align} 
Observe that $|A| \leq n$ since the first moment vanishes, i.e.,   $m_1 = 0$.  Now inserting \eqref{CdEB} in \eqref{0finalestonD} yields
\muun{
\label{DCdEB}
 &\left( \sum_{A \in \mathcal{A}_{2n} }  D_{n,A,L}[E + i \eta;\psi_1] \right)^{1/2} \nn \\
 &\leq \left( \sum_{A \in \mathcal{A}_{2n} }    \eta^{-|A|-1}   \| \psi_1 \|^2 \| \hat{B}_\# \|_{*,\infty}^{|A|}   [  C(E,d, L, \eta,\hat{B}_\#)   ]^{2n-|A|}   \left|  \prod_{a \in A} m_{|a|}  \right|  \right)^{1/2} \nn \\
 &\leq \| \psi_1 \| \left( \sum_{A \in \mathcal{A}_{2n} }  \eta^{-|A|-1}   \| B \|_1^{|A|}    \left|  \prod_{a \in A}  m_{|a|}  \right|   \left[   \tilde{C}(d,E,B) \left(1+ \ln \left(  \eta^{-1}  +1   \right)   \right) \right]^{2n-|A|}   \right)^{1/2} 
 \nn \\
&\leq  \| \psi_1 \|  \eta^{-(n+1)/2} \left( 1+ \ln \left(  \eta^{-1}  +1    \right)  \right)^{n}  \left( \sum_{A \in \mathcal{A}_{2n} }   \| B \|_1^{|A|}    \left|  \prod_{a \in A}  m_{|a|} \right|    \tilde{C}(d,E,B)  ^{2n-|A|}   \right)^{1/2}.
}
We define 
\begin{align} \label{defofKndEB} 
K_{n,d,E,B} :=\left( \sum_{A \in \mathcal{A}_{2n} }   \| B \|_1^{|A|}    \left|  \prod_{a \in A}  m_{|a|}  \right|    \tilde{C}(d,E,B)  ^{2n-|A|}   \right)^{1/2},
\end{align} 
where the notation does not reflect the dependence on the distribution ${\bf P}_v$ an hence the momenta $m_{|a|}$.
Thus  using  \eqref{eq:expansionU} -- \eqref{defofENL},  as well as \eqref{DCdEB} we find for the  LHS of  \eqref{eq:boundonexp} 
\begin{align*}
& \left|  \E \inn{ \psi_{1,\#} , (H_{  \lambda ,  L} - E  \mp  i  \eta  )^{-1}      \psi_{2,\#} }_{L^2(\Lambda_L)}  -  \sum_{j=0}^{n} T_{j,L}[E  \pm    i \eta  ; \psi_1, \psi_2]   \right| \\
 & \leq  \lambda^n  | U_{n,L}[z;\psi_1,\psi_2 ; \lambda ] |    \\
& \leq   \lambda^n  E_{n,L}[z;\psi_1]^{1/2}   \eta^{-1} \| \psi_{2,\#} \|_{L^2(\Lambda_L) }     \\
& \leq   \lambda^n \left( \sum_{A \in \mathcal{A}_{2n} }  D_{n,A,L}[z;\psi_1] \right)^{1/2}   \eta^{-1} \| \psi_{2,\#} \|_{L^2(\Lambda_L) }     
\\
& \leq  \lambda^n \eta^{-n/2} K_{n,d,E,B}  \left(  1 + \ln (  \eta^{-1} + 1 )  \right)^{n}  | \eta|^{-3/2}  \| \psi_1 \| \| \psi_2 \|    , 
\end{align*} 
where we used $\| \psi_{2,\#} \|_{L^2(\Lambda_L) }  \leq  \| \psi_2 \|  $, see Lemma \ref{fourierseries} (a).  
The uniform bound of  $K_{n,d,E,B}$ as a function of $E$ on compact subsets of the positive real line, follows directly from the 
explicit definition, c.f.  \eqref{defofKndEB},  \eqref{defoftildeCexplicit}, and  \eqref{C1}.  
This shows (a).  \\
Part  (b) follows for $\epsilon \in (0,2)$ by  inserting $\eta = \lambda^{2 - \epsilon}$ into (a).  Since then  we see that the RHS of 
\eqref{eq:boundonexp}
 tends to 
zero as $\lambda \to 0$ for $n$ sufficiently large. Specifically,  since logarithms are bounded by  any positive power, we see that    there exists a constant $C$ such that for  $\eta = \lambda^{2 - \epsilon}$ we have for small $\lambda$ 
\begin{align}
  \left( \frac{ \lambda^2  }{\eta}   \right)^{n/2} \left(1 +   \ln(   \eta^{-1} + 1 )\right)^n    \eta^{-{3/2}}   \leq C  \lambda^{  n - (2-\epsilon) \frac{n}{2}  -     n  \frac{ \epsilon}{4}  - (2 - \epsilon) \frac{3}{2}    } = C  \lambda^{\epsilon (\frac{n}{4}+\frac{3}{2}) -3}.
\end{align} 
Now  $ \lambda^{\epsilon (\frac{n}{4}+\frac{3}{2}) -3} \leq   \lambda^N$ provided  
$
 \epsilon (\frac{n}{4}+\frac{3}{2}) -3   \geq N  ,
$
i.e., 
$
 n    \geq 4  \frac{ N +3}{\epsilon }  - 6   .
$

\end{proof}

\label{proofofsec27}

\section{Acknowledgements}

D.H. wants to thank Laszlo Erd\"os for valuable discussions about Feynman graphs. 
We thank Robert Hesse, Benjamin Hinrichs,  and Oliver Siebert for helpful comments. 

%
%

\appendix
\section{Resolvent Expansions} \label{AppendixA}
In this chapter we want to display a very basic concept - the resolvent expansion - which we apply in Section 2.  We express the perturbed resolvent as a sum of products of unperturbed resovents and potential factors, which is the content of Lemma \ref{resolventenformel_interiert}.
To proof this expansion we begin with the following underlying resolvent identity. 
\begin{lemma}[Second resolvent identity]
\label{ResId}
Let $T,S$ be linear operators on the normed space $E$ with $D(S)=D(T)$. Let $z \in \rho (T) \cap \rho (S)$. Then we have
\muu{
\frac{1}{T-z}-\frac{1}{S-z}=\frac{1}{T-z}(S-T)\frac{1}{S-z}.
}
\end{lemma}
The following lemma is an immediate consequence.
\begin{lemma}
\label{Resolventen_Anwendung}
Let $A$ and $B$ be linear operators on a normed space $E$ with $D(A)=D(A + B)$ and $0 \in \rho (A) \cap \rho (A + B)$.
Then one has 
\muu{
\frac{1}{A+B} = \frac{1}{A} + \frac{1}{A+B}(-B)\frac{1}{A}.
}
\end{lemma}
\begin{proof}
By the second resolvent identity used for $T =A+B$, $S=A$ and $z=0$, we have
\muu{
\frac{1}{A+B}-\frac{1}{A}= \frac{1}{A+B}(A-(A+B))\frac{1}{A}  =  \frac{1}{A+B}(-B)\frac{1}{A}.
}
Now Lemma \ref{ResId} yields the statement.
\end{proof}
We can now perform the resolvent expansion using an iteration of the formula of the previous lemma.
\begin{lemma}
\label{resolventenformel_interiert}
Let $A$ and $B$ be linear operators on a normed space $E$ with $D(A)=D(A + B)$ and $0 \in \rho (A) \cap \rho (A + B)$.
Then for $m \in \N$ we have
\muu{
\frac{1}{A+B} = \sum_{n=0}^{m-1} \frac{1}{A} \left[ (-B) \frac{1}{A}  \right]^n  + \frac{1}{A+B} \left[ (-B) \frac{1}{A}  \right]^m .
}
\end{lemma}
\begin{proof}
We will prove the statement by induction.
For $m=0$
the sum is $0$ and only
\muu{
\underbrace{ \left[ \frac{1}{A} (-B) \right]^0 }_{=\operatorname{Id}} \frac{1}{A+B}=\frac{1}{A+B}
} 
remains.
Now let 
\muu{
\frac{1}{A+B} = \sum_{n=0}^{m-1}\frac{1}{A} \left[ (-B)\frac{1}{A}  \right]^n +\frac{1}{A+B}\left[ (-B)\frac{1}{A}  \right]^m 
}
hold for an $m \in \N$.
Using Lemma \ref{Resolventen_Anwendung} we have 
\muu{
\frac{1}{A+B} &= \sum_{n=0}^{m-1} \frac{1}{A} \left[(-B)  \frac{1}{A} \right]^n+ \left( \frac{1}{A} +\frac{1}{A+B} (-B) \frac{1}{A} \right) \left[  (-B)\frac{1}{A} \right]^m .
\\
&= \sum_{n=0}^{m-1} \frac{1}{A} \left[(-B) \frac{1}{A}  \right]^n  + \frac{1}{A} \left[ (-B) \frac{1}{A}  \right]^m+  \frac{1}{A+B} \left[(-B)\frac{1}{A} \right]^{m+1}
\\
&= \sum_{n=0}^{m}  \frac{1}{A}\left[ (-B) \frac{1}{A}  \right]^n + \frac{1}{A+B} \left[(-B) \frac{1}{A}  \right]^{m+1} .
}
\end{proof}

\section{Examples of dilation analytic Functions} 

\label{lemmanadilpotproof} 

In this appendix  we proof Lemma \ref{lem:anaprop} to give  a class of functions  which  satisfy    Hypotheses \ref{H1}  and  \ref{H2}, respectively. 
For this we recall the notation  
 $\langle x \rangle = (1 + x^2)^{1/2}$.

\begin{lemma}\label{lem:anaprop}  Let $x_0 , a \in \R^d$,  $\sigma > 0$,   $P$ be a polynomial on $\R^d$, and  $$f(x) = P(x) \exp(- \pi \sigma |x-x_0|^2  + 2 \pi i x \cdot a )$$  for $x \in \R^d . $
Then   $\theta \mapsto \hat{f}_\theta := (u(\theta) f)^\wedge$ - where $u(\theta)$ was defined in \eqref{equtheta} -
has an analytic extension from $\R$  to the strip $\{ z \in \C : | {\rm Im}  z| < \pi/4 \}$  as an $L^p(\R^d)$-valued function for all $p \in [1,\infty]$.  
\end{lemma} 
\begin{proof}
First observe that for the Gaussian on $\R^d$ given by  $g_\sigma(x) = \exp[-\pi \sigma |x|^2]$ where  $x \in \R^d$,  we have for $k \in \R^d$ that  $\hat{g_\sigma}(k) = \sigma^{-d/2} \exp[-\pi |k|^2/\sigma]$  (see for example \cite[Theorem 5.2]{LiebLoss.2001}). 
Thus,  using that the Fourier transform is 
bijective, that it  turns multiplication operators into differential operators, and translations into multiplication by a free wave function and vice versa \cite{LiebLoss.2001,folland},   we find 
 $\hat{f}(k) = Q_\sigma(k-a)   e^{-\frac{\pi}{\sigma} | k-a|^2} e^{ - 2 \pi  i (k-a)  \cdot x_0 } $ for some polynomial $Q_\sigma$. 
Hence multipliying out the expressions involving $k-a$ we find that there exists a polynomial $\tilde{Q}_{\sigma,a}$ and a complex vector $u \in \C^d$ depending on $a$ and $x_0$ such that 
$$
\hat{f}(k) = \tilde{Q}_{\sigma,a}(k)   e^{-\frac{\pi}{\sigma} | k|^2} e^{ - k \cdot u } 
$$
From   \eqref{eq:dilfour}  and the definition we see that 
 \begin{equation} \label{analyt1}
 \hat{f}_\theta(k) = e^{-d \theta/2} \hat{f}(e^{-\theta }k ) =  e^{-d \theta/2}  \tilde{Q}_{\sigma,a}(e^{-\theta }k )   e^{-\frac{\pi}{\sigma}  e^{-2\theta }|k|^2} e^{ - e^{-\theta} k \cdot u } .
 \end{equation} 
We have to show that  $ \theta \mapsto \hat{f}(e^\theta \cdot )$ has an analytic $L^p(\R^d)$-extention for $p \in [ 1,\infty ]$. \\
So, let  $p \in [1,\infty]$. 
Observe that   \eqref{analyt1} is for each fixed $k \in \R^d$ analytic in $\theta$ and the pointwise analytic extension satisfies the following bound. For $r > 0$ and $w \in \C$ let  
$D_r(w) = \{ z \in \C : |  z - w  | < r \}$. 
Fix  $\theta_0 \in (0,\pi/4)$ and $w \in \R$ and let $D = D_{\theta_0}(w)$.   Since $\theta_0 < \pi/4$ it follows that   
\muun{
\label{defdeltageq0}
\delta := \inf_{\theta \in D} {\rm Re} ( e^{-2  \theta}) > 0
}
holds.
 We  write \eqref{analyt1} as a product 
  \begin{align}  \label{anaprod1} 
    \hat{f}_\theta(k) = e^{-d \theta/2}   \left(   \tilde{Q}_{\sigma,a}(e^{-\theta }k )  e^{-\frac{\pi }{ \sigma } \frac{ \delta }{4} |k|^2} \right) \left(   e^{-\frac{\pi}{\sigma}  e^{-2\theta }|k|^2}  e^{ \frac{\pi }{\sigma} \frac{\delta}{ 2}  |k|^2} \right) \left(   e^{ -  e^{-\theta }k  \cdot u }  e^{- \frac{\pi }{\sigma}\frac{\delta}{4}  |k|^2}  \right)  . 
 \end{align} 
To estimate the middle factor in brackets we  observe  that by the choice of $\delta$
 \begin{align} \label{anaprod2}
-  {\rm Re} \,  e^{-2 \theta} |k|^2 +   \frac{\delta}{2}  |k|^2 \leq  -   \frac{\delta}{2}  |k|^2  \quad \text{for all }  k \in \R^d  , \theta \in D .
 \end{align} 
 Using \eqref{anaprod1}, we see from \eqref{anaprod2} and the Gaussians in the first and third factor that there exists    a constant $C$ such that  
 for all $\theta \in D$,  $n=0,1$ and $k \in \R^d$   we have 
\begin{equation} \label{eq:parder}
|\partial_\theta^{n}  \hat{f}_\theta(k) |  \leq C \exp \left(- \frac{\pi}{\sigma} \frac{ \delta}{2}  |k|^2 \right).
\end{equation}
Now let  $q \in [1,\infty]$ with  $\frac{1}{p} + \frac{1}{q} = 1$.  From \eqref{eq:parder}
 and dominated convergence   we see that  for any $h \in L^q(\R^d)$ the function $\theta \mapsto \int h(k) f_\theta(k) dk$
 is analytic (explicitly use for example  \cite[Theorem 2.27]{folland} and the Cauchy-Riemann equations).
 For $p \in [1,\infty)$ we have  the duality relation   $(L^p(\R^d))^* \cong L^q(\R^d)$,  cf. \cite[Theorem 2.45]{AdamsFournier.2002}. Thus for $p \in [1,\infty)$, it follows  that  $\theta \mapsto \hat{f}_\theta$ is 
weakly analytic  in $L^p(\R^d)$, and therefore, 
it is in fact strongly analytic in $L^p(\R^d)$ by  the well known fact that weak analyticity in Banach spaces implies 
strong analyticity,  cf.  \cite[Theorem VI.4]{SR1}.  Hence we have shown the claim for $p \in [1,\infty)$.  \\
 To prove  analyticity in  $L^\infty(\R^d)$, we need a  different argument, since the dual of $L^\infty(\R^n)$ is larger than  $L^1(\R^d)$. 
First we show  that   $\hat{f}_\theta $   is a continuous  $L^\infty(\R^n)$-valued function of $\theta \in D$ by 
showing this   for  each factor  in \eqref{anaprod1}. 
For  $\theta \in D$ and $w = e^{-\theta}$   
 we see from  the mean value   theorem  that there exists  a  constant $C$ and an $m \in \N_0$ such that for all $k \in \R^d$ and $h \in \C$
\begin{align} \label{cont1}
 \left|    \tilde{Q}_{\sigma,a}((w + h )k )   -  \tilde{Q}_{\sigma,a}(w k  )\right|e^{-\frac{\pi }{ \sigma } \frac{ \delta }{4} |k|^2} \leq  |h | C \langle k   \rangle^m e^{-\frac{\pi }{ \sigma } \frac{ \delta }{4} |k|^2}  ,
\end{align}  
\begin{align} \label{cont2}
 \left|  e^{-(w+h) k \cdot u} -  e^{-w k \cdot u}   \right| e^{-\frac{\pi }{ \sigma } \frac{ \delta }{4} |k|^2} \leq  |h | C  |k| \sup_{z: |z| \leq |h|} | e^{-  (w+z ) k \cdot u }|  e^{-\frac{\pi }{ \sigma } \frac{ \delta }{4} |k|^2} , 
\end{align} 
\begin{align} 
 \left|   e^{-\frac{\pi}{\sigma}( w + h)^2 |k|^2}   - e^{-\frac{\pi}{\sigma} w^2 |k|^2}  \right| e^{ \frac{\pi }{\sigma} \frac{\delta}{ 2}  |k|^2} 
 &  \leq  |h | C  |k|^2 \sup_{z: |z| \leq |h|} \left| (w+z) e^{-\frac{\pi}{\sigma} (w+z )^2|k|^2} \right|  e^{ \frac{\pi }{\sigma} \frac{\delta}{ 2}  |k|^2} \nn \\
 &   \leq  |h | \tilde{C}  |k|^2  e^{- \frac{\pi }{\sigma} \frac{\delta}{ 4}  |k|^2} , \label{cont3}
\end{align} 
where the last inequality in \eqref{cont3}  holds  for  some constant $\tilde{C}$ provided 
 $|h|$ is  sufficiently small,  since by   \eqref{defdeltageq0}  and   continuity  there exists an $r > 0$  such that for all 
 $\theta \in D$  and $z \in \C$ with $|z| < r$  we have 
$ {\rm Re} ( e^\theta + z)^2 \geq  \frac{3}{4} \delta$.
Thus $L^\infty$--continuity of $\theta \mapsto \hat{f}_\theta $ now follows from  \eqref{anaprod1}  and \eqref{cont1}--\eqref{cont3} and the fact that the 
exponential map $\theta \mapsto  e^{-\theta} =:w $  is a continuous function on $\C$.
Now let $\gamma$ be a closed pieceweise $C^1$--curve   in $D$. 
Then  we  define 
 $
 \int_\gamma   \hat{f}_\theta d\theta 
 $
 as a Riemann integral in $L^\infty(\R^d)$, which exists by continuity of the integrand as a function of $\theta$.
We want to show that this integral vanishes. For this, 
we introduce the bilinear form $\llangle  \cdot , \cdot \rrangle   : L^1(\R^d) \times L^\infty(\R^d)\to \C$,  $(f,g) \mapsto  \int_{\R^d}  f(k) g(k) dk$.
This bilinear form is continuous, since an elementary estimate shows that  $|\llangle   f ,  g  \rrangle | \leq \| f \|_1 \| g \|_\infty$. 
Let $h \in L^1(\R^d)$  be arbitrary. By what we have shown above 
we know  that  $\theta \mapsto    \llangle    h ,  \hat{f}_\theta  \rrangle  =  \int h(k) f_\theta(k) dk$  is analytic. 
Thus it follows from  continuity of  $\llangle  \cdot , \cdot \rrangle$ that
 \begin{align} \label{identmorera} 
 \llangle    h ,  \int_\gamma   \hat{f}_\theta d\theta  \rrangle    =  \int_\gamma  \llangle    h ,  \hat{f}_\theta  \rrangle   d\theta = 0 .  
 \end{align} 
 Using $(L^1)^* \cong L^\infty$, we see from  \eqref{identmorera} since  $h \in L^1(\R^d)$ was arbitrary  that  $\int_\gamma   \hat{f}_\theta d\theta  = 0$. 
 Since this holds for any closed piecewise $C^1$--curve $\gamma$ in $D$, strong analyticty of 
$\theta \mapsto \hat{f}_\theta$ in $L^\infty(\R^d)$ now  follows  from a straightforward  generalization of 
 Morera's theorem to Banach space valued functions, see for example  \cite[1.5, Exercise 2]{AbramovichAliprantis.2002}.

\end{proof}

\section{Estimates on Integrals} 
\label{AppendixC}

In this Appendix, we present a collection of estimates on integrals as well as on discrete integrals, i.e.,  infinite sums.

Let us first mention a basic inequality.  For any $z \in \C$ we have the elementary bound 
\begin{equation}  \label{ln-1}
   | z^{-1}|   \leq   \left|   {\rm Im} z \right|^{-1}  .
\end{equation} 
The following lemma is a simple estimate,  which is applied in the proof of Lemma \ref{lemestfourdisc}.
\begin{lemma}
\label{ChSymbEst}
For  $a,b \in \R$ with $a \leq b$ and  $ \langle \cdot \rangle^2 f \in L^\infty$ with $\langle \cdot \rangle$ defined in \eqref{Klammern}.  Then  $f \in L^1(\R)$ and
 \begin{align*}
\int_{a}^b  \left|   f(x) \right|   dx \leq \pi \sup_{x \in \R} | \langle x \rangle^2 f(x) | .
\end{align*}
\end{lemma}
\begin{proof}
Using an estimate on the $\arctan$ and a straight forward calculation shows
 \begin{align*} 
\int_{a}^b  \left|   f(x) \right|   dx
 &\leq    \int_{a}^{b} \langle x \rangle^{-2}\langle x \rangle^{2}  |  f(x) |  dx \leq  \sup_{x \in \R} | \langle x \rangle^2 f(x) |   \int_{a}^{b}\frac{1}{1+x^2} dx \\ &  =  \sup_{x \in \R} | \langle x \rangle^2 f(x) | \underbrace{\left[ \arctan(x) \right]_a^b}_{\leq \pi}  \leq \pi \sup_{x \in \R} | \langle x \rangle^2 f(x) | < \infty .
\end{align*}
\end{proof}
Moreover we shall need to estimate tracial expressions of resolvents. 
Let us now  estimate a  finite volume expression. For this we shall make use of the following elementary lemma about Riemann integrals.

\begin{lemma}\label{lem:discrriem}  Let $I  = [-c,c]^d$ and $I_j$, $j=1,...,N$, be   a partition (up to boundaries) of $I$  into translates of a square  $Q$ which is centered at the 
origin. Let $\xi_j  \in  I_j$
and $ \Delta x_j = | I_j|$. Suppose $g \geq 0$ is a Riemann integrable function on $I$  and we have 
\begin{equation} \label{eq:estonint} 
\sup_{ \xi \in I_j} f(\xi) \leq \inf_{ \xi \in I_j} g(\xi)  . 
\end{equation} 
Then 
\begin{align*} 
  \left|  \sum_{i=1}^N f(\xi_j) \Delta x_j  \right| \leq  \int_I g(x) dx  .
\end{align*} 
\end{lemma}
\begin{proof}  The statement follows directly from the theory of Riemann integration. 
\end{proof}

We now want to show Proposition \ref{C007} which is used in Chapter \ref{sec:AsymEE} to estimate integrals over resolvent-type-functions. To do so, we are first discussing the relation between the discrete sum and the Riemann integral in Lemma \ref{ln0}, to then give a bound on the Riemann integral in Lemma \ref{ln00},  with which we can then show Proposition \ref{C007}.

\begin{lemma} Let $f : \Lambda_L^* \to \C$ and $E \geq  0$. Then 
\label{ln0}
$$
\int_{\Lambda_L^*}   \left|  \frac{  f(q) }{  q^2 - E -  i \eta } \right|    dq 
  \leq C_0(E,d, L , \eta,f)  
$$
where 
\begin{align} \label{C0}
C_0(E,d, L, \eta,f) := \| f \|_{*,\infty} \int\limits_{ \| q \|_\infty \leq  \sqrt{ 2  E  + 1  }  }   |q^2 - E - i 2^{-1/2}  \eta  |^{-1}   dq  + (E +1 )^{-1} \| f \|_{*,1} .
\end{align}
where we introduced the notation $\| q \|_\infty = \max_{j=1}^d |q_j|$. 
\end{lemma}

\begin{proof}  To prove the Lemma we want to use Lemma \ref{lem:discrriem}. 
For this we  need  to verify  \eqref{eq:estonint}.
Let $q,\xi,\xi_1,\xi_2 \in \R^d$, $s \geq 1$,  and $|q|^2 \leq s  E$.  First observe that 
\begin{align*} 
  |(q+\xi)^2 - E - i \eta |^2    & = | q^2 - E + \xi^2 + \xi q + q \xi - i \eta  |^2  \\
  & = (q^2 - E )^2 + 2 (q^2 - E)  ( \xi q + q \xi + \xi^2 )  +  ( \xi q + q \xi + \xi^2 )^2 + \eta^2. 
\end{align*} 
From this  we find for   $d_j =  \xi_j q + q \xi_j + \xi_j^2$ that 
\begin{align}
 &  |(q+\xi_1)^2 - E - i \eta_1 |^2  -   |(q+\xi_2)^2 - E - i \eta_2 |^2  \nonumber \\
  & =  2 (q^2 - E)  ( d_1 - d_2 )  +  d_1^2 - d_2^2 +  \eta_1^2 - \eta_2^2   \geq  \frac{1}{4} \eta^2 \geq 0, \label{eq:inproofofsomelemma} 
\end{align} 
provided $\sqrt{2} \eta_2 =  \eta_1 = \eta $ and 
\begin{equation} \label{eq:estond} 
| d_1 | , |d_2 | \leq \min \left\{ \eta \frac{1}{4} , \eta^2 \frac{1}{16}   \frac{1}{  2 s  E } \right\}   =: C(s,\eta). 
\end{equation} 
Thus if 
$$
| \xi_j | \leq \max \left\{1,\frac{ C(s,\eta)}{2 \sqrt{s E} + 1 }  \right\} =: D(s,\eta) , 
$$
then \eqref{eq:estond}  holds,   and hence for all $q$ with $|q| \leq \sqrt{ s E}$ we find from  \eqref{eq:inproofofsomelemma} 
\begin{align*} 
&  \frac{1}{ |(q+\xi_1)^2 - E - i \eta |}  \\
&  = \frac{1}{| (q+\xi_2)^2 - E - i  2^{-1/2} \eta |  +  |(q+\xi_1)^2 - E - i \eta |- | (q+\xi_2)^2 - E - i  2^{-1/2} \eta |     }     \\
& \leq \frac{1}{  |(q+\xi_2)^2 - E - i 2^{-1/2}  \eta | }  . 
\end{align*} 
This implies 
\begin{align*} 
& \sup_{\xi : |\xi| \leq D(s,\eta) }  |(q+\xi)^2 - E - i \eta |^{-1}  
 \leq \inf_{\xi : |\xi| \leq D(s,\eta) }  |(q+\xi)^2 - E - i 2^{-1/2}  \eta |^{-1}  . 
\end{align*} 
Hence  \eqref{eq:estonint} is satisfied and we find for $L > \frac{ \sqrt{d/2} }{  D(s,\eta)}$ that 
$$
  \left| \int_{\Lambda_L^*}   \left( q^2 - E -  i \eta \right)^{-1}  1_{ | q_j | \leq  \sqrt{ s E / d }  } dq \right|   
  \leq\int   |q^2 - E - i 2^{-1/2}  \eta |^{-1} 1_{ | q_j | \leq  \sqrt{ s E / d }  }  dq . 
$$
Now for $s  =  d (2 + E^{-1}) $ we have 
\begin{align*}
&   \int_{\Lambda_L^*}  \left| f(q)  \left( q^2 - E -  i \eta \right)^{-1}   \right|  dq  \\
 & \leq    \int_{\Lambda_L^*}  \left| f(q) \left( q^2 - E -  i \eta \right)^{-1} \right|   1_{ | q_j | \leq    \sqrt{ s E / d }  } dq +   \int_{\Lambda_L^*}  \left|  f(q)  \left( q^2 - E -  i \eta \right)^{-1} \right|  1_{ | q_j |  >   \sqrt{ s E / d }  }  dq  \\
 &\leq \| f \|_{*,\infty} \int_{\{ q \in \R^d :  \|  q \|_\infty \leq  \sqrt{ 2  E  + 1  } \}  }   |q^2 - E - i 2^{-1/2}  \eta  |^{-1}   dq  + (E +1 )^{-1}  \int_{\Lambda_L^*} |f (q)  | dq, 
\end{align*} 
which completes the proof.  
\end{proof}

\begin{lemma} 
\label{ln00}
For $f \in L^1 (\R^d) \cap L^\infty (\R^d)$,  $E >0 $,  $\eta >0$ and $d \in \N$ we have
\begin{align*} 
    \left| \int_{\R^d} \frac{ f(q) }{ q^2 - E \pm  i \eta }  dq \right| 
      \leq  C_1(E,d)  \| f \|_\infty    \ln \left(  \frac{1}{ \eta }  +1    \right) + \sqrt{2} \| f \|_1,
\end{align*}
where 
\begin{align} \label{C1}
C_1(E,d):=  \sqrt{2}  \left(  E^{-1/2} ( E+1)^{\frac{d-1}{2}}  + E^{\frac{d-2}{2}} \right)  |S_{d-1}|, 
\end{align}
 with  $|S_{d-1}|$  being the volume of the $d-1$-dimensional sphere with radius $1$.

\end{lemma}

\begin{proof}
Let $\delta >   0$ be arbitrary.  Using $ \sqrt{ 2 ( |x|^2 + |y|^2 ) } \geq ( |x| + |y| ) $ we find 
\begin{align}
   \left| \int_{\R^d}  \frac{ f(q) }{ q^2 - E \mp   i \eta }  dq \right|  \label{WAFest1} 
& \leq    \int_{\R^d}  \frac{\sqrt{2} | f(q) |}{| q^2 - E| + \eta}   dq   \\
& =     \int_{\R^d}   1_{|q^2 - E| \leq \delta} \frac{\sqrt{2} | f(q) | }{| q^2 - E| + \eta}   dq +   \int_{\R^d}   1_{|q^2 - E| >  \delta}   \frac{\sqrt{2}| f(q) |}{| q^2 - E| + \eta}   dq.  \nn 
\end{align} 
We will now consider both summands separately.
For the first one we estimate 
\begin{align} \label{WAFest2}
 &    \int_{\R^d}   1_{|q^2 - E| >  \delta}  \frac{\sqrt{2} | f(q) |}{| q^2 - E| + \eta}   dq  \leq \frac{\sqrt{2}\| f\|_1}{\delta}.
\end{align} 
Independent of the dimension we can perform the following estimate.
We are going over to $d$-dimensional polar coordinates and resolve the absolute value in the condition of the indicator function.  Let $a_+ : = \max\{a,0\}$. 
We write  
\begin{align}
\label{IIestonq}  
   &  \int_{\R^d}   1_{|q^2 - E| \leq \delta}  \frac{\sqrt{2} | f(q) |}{| q^2 - E| + \eta}   dq  \nn \\
& =      \int_0^\infty \int_{S_{d-1}} 1_{|r^2 - E| \leq   \delta}  \frac{\sqrt{2} | f(r \omega ) |}{| r^2 - E| + \eta} d\omega   r^{d-1} dr   \nn   \\
& =       \int_0^\infty \int_{S_{d-1}} 1_{ 0 \leq  r^2 - E  \leq   \delta}  \frac{\sqrt{2} | f(r \omega ) |}{| r^2 - E| + \eta} d\omega   r^{d-1} dr    +    \int_0^\infty \int_{S_{d-1}} 1_{ 0 \leq  E - r^2  \leq   \delta}   \frac{\sqrt{2}| f(r \omega ) |}{| r^2 - E| + \eta} d\omega   r^{d-1} dr   \nn \\
& =     \underbrace{    \int_{S_{d-1}} \int_{\sqrt{E}}^{ \sqrt{E+\delta}}   \frac{\sqrt{2} | f(r \omega ) |}{| r^2 - E| + \eta} d\omega   r^{d-1} dr   }_{=: I_1}  +  \underbrace{  \int_{S_{d-1}}  \int_{\sqrt{(E - \delta)_+} }^{\sqrt{E}} \frac{\sqrt{2} | f(r \omega ) | }{| r^2 - E| + \eta} d\omega   r^{d-1} dr  }_{=: I_2 }.
\end{align}  
For these two summands, we can use  $r^2 - E = (r-\sqrt{E})(r + \sqrt{E})$  in the denominator to get
\begin{align} 
\label{i1start}
I_1 & \leq     \| f \|_\infty    |S_{d-1}| \int_{\sqrt{E}}^{ \sqrt{E+\delta}}  \frac{\sqrt{2}}{(r -\sqrt{E}) (r +\sqrt{E} )  + \eta}    r^{d-1} dr    
\end{align}
as well as
\begin{align} 
\label{i2start}
I_2 & \leq     \| f \|_\infty    |S_{d-1}|  
\int_{\sqrt{(E - \delta)_+} }^{\sqrt{E}}   \frac{\sqrt{2}}{(\sqrt{E}-r)(\sqrt{E} +r)  + \eta}   r^{d-1} dr. 
\end{align}
Now using the trivial estimate   $\frac{1}{r+\sqrt{E}} \leq \frac{1}{\sqrt{E}} $ lets us perform the following estimates for the integral in \eqref{i1start}
\begin{align} 
\label{i1integral}
\int_{\sqrt{E}}^{ \sqrt{E+\delta}}  \frac{\sqrt{2}}{(r -\sqrt{E}) (r +\sqrt{E} )  + \eta}    r^{d-1} dr      
&\leq     \int_{\sqrt{E}}^{ \sqrt{E+\delta}}  \frac{\sqrt{2}}{(r -\sqrt{E})\sqrt{E}  + \eta}    r^{d-1} dr  \nn
\\
& \leq     \frac{1}{E^{\frac{1}{2}}} \int_{\sqrt{E}}^{ \sqrt{E+\delta}}  \frac{\sqrt{2}}{ r -\sqrt{E} + \eta E^{-1/2}}    (E + \delta)^{\frac{d-1}{2}} dr    \nn
\\
& =    \sqrt{2}    \frac{( E+\delta)^{\frac{d-1}{2}}}{E^{\frac{1}{2}}} \left[  \ln \left( r - \sqrt{E} + \eta E^{-1/2} \right) \right]_{\sqrt{E}}^{ \sqrt{E+\delta}},
\end{align}
as well as for the one in \eqref{i2start}
\muun{
\label{i2integral} 
\int_{\sqrt{(E - \delta)_+} }^{\sqrt{E}}   \frac{\sqrt{2}}{(\sqrt{E}-r)(\sqrt{E} +r)  + \eta}   r^{d-1} dr 
&  \leq     
  \int_{\sqrt{(E - \delta)_+} }^{\sqrt{E}}   \frac{\sqrt{2}}{(\sqrt{E}-r)\sqrt{E}  + \eta}   r^{d-1} dr   
\nn
  \\
& \leq   
 \frac{1}{E^{1/2}}   \int_{\sqrt{(E-\delta)_+}}^{\sqrt{E}}   \frac{\sqrt{2}}{ \sqrt{E}-r + \eta E^{-1/2}}   E^{\frac{d-1}{2}} dr  \nn
  \\
& =    
   E^{\frac{d-2}{2}}  \left[ \ln \left( \sqrt{E} - r  + \eta E^{-1/2} \right) \right]_{ \sqrt{(E-\delta)_+} }^{ \sqrt{E}}.   
}

We can finally evaluate the logarithm-terms in the RHS of \eqref{i1integral} via
\begin{align}
\label{i1ln}
 \left[  \ln \left( r - \sqrt{E} + \eta E^{-1/2} \right) \right]_{\sqrt{E}}^{ \sqrt{E+\delta}}
 & =      \ln \left( \frac{ \sqrt{E+\delta} - \sqrt{E} +   \eta E^{-\frac{1}{2}}}{  \eta E^{-\frac{1}{2}}  }  \right)  \nn
    \\
 & =   \ln \left(  \frac{ \sqrt{E+\delta} - \sqrt{E}}{ \eta E^{-1/2}} + 1 \right)  \nn
    \\
& =  \ln \left(  \frac{ (E+\delta) - E }{ \eta  (\sqrt{1+E^{-1}\delta} +1) } + 1 \right) \nn
 \\
   & \leq   \ln \left(  \frac{ \delta }{ \eta} + 1 \right),
\end{align} 
as well as the one in \eqref{i2integral} 
\muun{ 
\label{i2ln}
 \left[ \ln \left( \sqrt{E} - r  + \eta E^{-1/2} \right) \right]_{ \sqrt{(E-\delta)_+} }^{ \sqrt{E}}    \nn
&=     
 \ln \left( \frac{\sqrt{E}  -  \sqrt{(E-\delta)_+} +   \eta E^{-\frac{1}{2}}}{  \eta E^{-\frac{1}{2}}}     \right)    \nn
    \\
 & =       \ln \left(  \frac{\sqrt{E} -  \sqrt{(E-\delta)_+} }{ \eta E^{-1/2} }  +1    \right)    \nn
    \\
      & =      \ln \left(  \frac{ E -  (E-\delta)_+ } {\eta(  1 + \sqrt{ ( 1 - E^{-1}  \delta )_+} )  }  +1    \right)    \nn
     \\
   & \leq      \ln \left(  \frac{\min\{\delta,E\}}{ \eta }  +1    \right) .
}
Putting together \eqref{i1start},  \eqref{i1integral} and \eqref{i1ln}  gives us
\muun{
\label{i1final}
I_1 \leq     \| f \|_\infty    |S_{d-1}| \sqrt{2}   \frac{( E+\delta)^{\frac{d-1}{2}}}{E^{\frac{1}{2}}}  \ln \left(  \frac{ \delta }{ \eta} + 1 \right),  
}
while  \eqref{i2start},  \eqref{i2integral} and \eqref{i2ln}  gives
\muun{
\label{i2final}
I_2 \leq        \| f \|_\infty    |S_{d-1}| \sqrt{2}   
  E^{\frac{d-2}{2}}  \ln \left(  \frac{\min\{\delta,E\}}{ \eta }  +1    \right) . 
}
Adding \eqref{i1final} and \eqref{i2final},  while using $\min\{ \delta, E \} \leq \delta$  we find from  \eqref{IIestonq}  that    
\begin{align} 
 \int_{\R^d}   1_{|q^2 - E| \leq \delta}  \frac{\sqrt{2} | f(q) |}{| q^2 - E| + \eta}   dq  = I_1 + I_2    & \leq     \left(   \frac{( E+\delta)^{\frac{d-1}{2}}}{E^{\frac{1}{2}}} 
   + 
    E^{\frac{d-2}{2}} \right) \| f \|_\infty    |S_{d-1}| \sqrt{2}  \ln \left(  \frac{\delta}{ \eta }  +1    \right). \label{WAFest4}
\end{align} 
Now  choosing  $\delta =1 $   
we find with  \eqref{WAFest1},  \eqref{WAFest2}, and  \eqref{WAFest4} that 
\begin{align*}
  &  \left| \int_{\R^d}  f(q) \left( q^2 - E \mp  i \eta \right)^{-1}  dq \right| \\
& \leq     \int_{\R^d}   1_{|q^2 - E| \leq \delta}  \frac{\sqrt{2} | f(q) |}{| q^2 - E| + \eta}   dq +   \int_{\R^d}   1_{|q^2 - E| >  \delta}  \frac{\sqrt{2} | f(q) |}{| q^2 - E| + \eta}   dq  \\
&\leq  C_1(E,d)  \| f \|_\infty    \ln \left(  \frac{1}{ \eta }  +1    \right) + \sqrt{2} \| f \|_1
\end{align*}
\end{proof}

The following proposition will be needed in Section \ref{sec:AsymEE}.

\begin{proposition}  
\label{ln000}
Let $E > 0$ and $\eta > 0$. Then for all $L \geq 1$ and $f : \Lambda_L^* \to \C$ we have  

$$
  \left| \int_{\Lambda_L^*}  f(q)  \left( \frac{1}{2} q^2 - E \pm  i \eta \right)^{-1}  dq \right|   
  \leq C(E,d, L , \eta,f) 
$$
where 
\begin{align} \label{C007}
C(E,d, L , \eta,f) :=  2 \| f \|_{*,\infty}  \left[  C_{1}(2 E,d)  \ln \left(  \eta^{-1}  +1    \right)+  2^d \sqrt{2} (4 E + 1)^{d/2} \right] + 2  \| f \|_{*,1} 
\end{align}
with $C_1(E,d)$ defined in  \eqref{C1}. 
\end{proposition} 
\begin{proof}
This follows by  combining Lemmas  \ref{ln0} and  \ref{ln00}.  In more detail, it follows from Lemma  \ref{ln0} that 
\begin{align} \label{eq:maindiscest1} 
  \left| \int_{\Lambda_L^*}  \frac{f(q) dq}{  \sfrac{1}{2} q^2 - E \pm  i \eta } \right|   
  \leq 2 \| f \|_{*,\infty} \int\limits_{ \{ q \in \R^d :\| q \|_\infty \leq  \sqrt{ 4  E  + 1  }  \} }   \frac{dq}{|q^2 - 2 E - i 2^{1/2}  \eta  |}     +\frac{2\| f \|_{*,1}}{2E +1 } , 
\end{align} 
recalling $\| q \|_\infty = \max_{j=1}^d |q_j|$.
Now using Lemma   \ref{ln00} to estimate the integral on the right hand side of  \eqref{eq:maindiscest1}, we find 
\begin{align} \label{eq:maindiscest2} 
\int\limits_{ \{ q \in \R^d :\| q \|_\infty \leq  \sqrt{ 4  E  + 1  }  \} }  \frac{dq}{|q^2 - 2 E - i 2^{1/2}  \eta  |} \leq  C_1(2 E, d) \ln\left( 2^{-1/2} \eta^{-1} + 1 \right)   +  2^d \sqrt{2} (4E+1)^{d/2} .
\end{align} 
Thus inserting  \eqref{eq:maindiscest2}  into \eqref{eq:maindiscest1} the inequality of the lemma follows using that by  montonicity of $\ln$ we have 
$ \ln\left( 2^{-1/2} \eta^{-1} + 1 \right) \leq \ln\left(\eta^{-1} + 1 \right) $. 
\end{proof}


\bibliography{Quellen}
\bibliographystyle{plain}


\end{document}